\documentclass[11pt, draftclsnofoot, onecolumn]{IEEEtran}

\usepackage{pifont}
\usepackage{cite}
\usepackage[dvips]{epsfig}
\usepackage{graphicx}
\usepackage{calligra}
\usepackage[T1]{fontenc}
\usepackage{bbold}
\usepackage{mathrsfs}

\usepackage{subfigure}
\usepackage{color}
\usepackage{amsmath}
\usepackage{cases}

\usepackage{amssymb}
\usepackage[justification=centering]{caption}
\usepackage{framed}

\usepackage{subfigure}
\usepackage{color}
\usepackage{amsmath}
\usepackage{bm}
\usepackage{amssymb}
\usepackage{amsbsy}
\usepackage{bbm}
\usepackage{dsfont}
\usepackage[normalem]{ulem}
\usepackage{xcolor,cancel}
\usepackage{enumerate}
\usepackage{paralist}

\usepackage{algorithm, algorithmic}

\usepackage{amsthm}

\newtheorem{theorem}{Theorem}

\newtheorem{corollary}{Corollary}

\newtheorem{definition}{Definition}

\newtheorem{proposition}{Proposition}

\newtheorem{lemma}{Lemma}

\newtheorem{remark}{Remark}

\newtheorem{assumption}{Assumption}

\newcommand{\bbE}{\mathbbm{E}}
\newcommand{\tbE}{\tilde{\mathbbm{E}}}

\newcommand{\tbP}{\tilde{\mathbbm{P}}}
\newcommand{\Ind}{\mathbbm{1}}

\newcommand{\sT}{\mathsf{T}}
\newcommand{\sD}{\mathsf{D}}

\newcommand{\teE}{\text{E}}
\newcommand{\teL}{\text{L}}

\newcommand{\cD}{\mathcal{D}}

\newcommand{\cE}{\mathcal{E}}
\newcommand{\cH}{\mathcal{H}}

\newcommand{\cN}{\mathcal{N}}
\newcommand{\cQ}{\mathcal{Q}}

\newcommand{\bone}{\mathbbm{1}}

\newcommand{\bPsi}{{\boldsymbol{\Psi}}}

\newcommand{\zo}{\zeta_{\rm{obj}}}

\newcommand{\tu}{\tilde{u}}
\newcommand{\tp}{\tilde{p}}
\newcommand{\tq}{\tilde{q}}
\newcommand{\tl}{\tilde{l}}
\newcommand{\tL}{\tilde{L}}

\newcommand{\ignore}[1]{{}}

\newcommand{\lp}{\left(}
\newcommand{\rp}{\right)}

\newcommand{\lsb}{\left[}
\newcommand{\rsb}{\right]}

\newcommand{\lcb}{\left\{}
\newcommand{\rcb}{\right\}}

\newcommand{\lbar}{\left|}
\newcommand{\rclose}{\right.}

\newcommand{\TL}{{\mathsf{T}}_{\rm{L}}}
\newcommand{\TE}{{\mathsf{T}}_{\rm{E}}}

\newcommand{\hTL}{{\hat{\mathsf{T}}}_{\rm{L}}}
\newcommand{\hTE}{{\hat{\mathsf{T}}}_{\rm{E}}}

\newcommand{\hTLO}{{\hat{\mathsf{T}}}_{\rm{L}}^{(1)}(\bPsi)}
\newcommand{\hTEO}{{\hat{\mathsf{T}}}_{\rm{E}}^{(1)}(\bPsi)}

\newcommand{\hTLZ}{{\hat{\mathsf{T}}}_{\rm{L}}^{(0)}(\bPsi)}
\newcommand{\hTEZ}{{\hat{\mathsf{T}}}_{\rm{E}}^{(0)}(\bPsi)}

\newcommand{\hTLi}{{\hat{\mathsf{T}}}_{\rm{L}}^{(i)}(\bPsi)}
\newcommand{\hTEi}{{\hat{\mathsf{T}}}_{\rm{E}}^{(i)}(\bPsi)}

\newcommand{\hlambda}{{\hat{{\lambda}}}}

\newcommand{\AL}{A_{\rm{L}}}
\newcommand{\AEE}{A_{\rm{E}}}
\newcommand{\BL}{B_{\rm{L}}}
\newcommand{\BE}{B_{\rm{E}}}

\newcounter{parentalgorithm}

\begin{document}

\title{Asymptotically Optimal Stochastic Encryption for Quantized Sequential Detection in the Presence of Eavesdroppers
%
}
%
%
%

\author{Jiangfan Zhang,~\IEEEmembership{Member,~IEEE},
	and Xiaodong Wang,~\IEEEmembership{Fellow,~IEEE}
%
%
%
%
%
%
}

\maketitle

\begin{abstract}
We consider sequential detection based on quantized data in the presence of eavesdropper. Stochastic encryption is employed as a counter measure that flips the quantization bits at each sensor according to certain probabilities, and the flipping probabilities are only known to the legitimate fusion center (LFC) but not the eavesdropping fusion center (EFC).
As a result, the LFC employs the optimal sequential probability ratio test (SPRT) for sequential detection whereas the EFC employs a mismatched SPRT (MSPRT).  We characterize the asymptotic performance of the MSPRT in terms of the expected sample size as a function of the vanishing error probabilities.  We show that when the detection error probabilities are set to be the same at the LFC and EFC, every symmetric stochastic encryption is ineffective in the sense that it leads to the same expected sample size at the LFC and EFC. 
Next, in the asymptotic regime of small detection error probabilities,  we show that every stochastic encryption degrades the performance of the quantized sequential detection at the LFC by increasing the expected sample size, and the expected sample size required at the EFC is no fewer than that is required at the LFC. Then the optimal stochastic encryption is investigated in the sense of maximizing the difference between the expected sample sizes required at the EFC and LFC. Although this optimization problem is nonconvex, we show that
 if the acceptable tolerance of the increase in the expected sample size at the LFC induced by the stochastic encryption is small enough, then the globally optimal stochastic encryption can be analytically obtained; and moreover, the optimal scheme only flips one type of quantized bits (i.e., $1$ or $0$) and keeps the other type unchanged. 
\end{abstract}


\section{Introduction}


Decentralized detection in sensor networks using quantized data has been extensively studied, see \cite{viswanathan1997distributed, blum1997distributed, chen2006channel, hashemi1989decentralized, veeravalli1993decentralized, veeravalli1999sequential, mei2008asymptotic, cheng2005bandwidth, fellouris2011decentralized, yilmaz2013channel, li2015multi, li2015decentralized} and references therein.
%
The focus of this paper is on the sequential hypothesis testing in a decentralized sensor network
based on quantized sensor data.
Specifically, each sensor sequentially takes samples
and then sends the binary quantized version of each sample to a legitimate fusion center (LFC). 
The LFC 
performs the sequential probability ratio test (SPRT) which is the optimal procedure for sequential detection of binary hypotheses in the sense of minimizing the expected sample sizes required for achieving the prescribed detection accuracy \cite{wald1947sequential, wald1948optimum}.

Due to the broadcast nature of the communication links, the communications between sensors and the LFC are inherently vulnerable to eavesdropping, and hence, sensor networks are susceptible to security breach, which is an important problem especially when the network is part of a larger cyber-physical system \cite{Poor03012017, kailkhura2015distributed, mukherjee2015physical}. For instance, some nodes within a cognitive radio (CR) network may eavesdrop on the transmissions from other nodes to the LFC, detect the vacant primary user channels, and encroach on the vacant primary user channels without paying any participation costs to the network moderator \cite{kailkhura2015distributed}. 
Eavesdropping fusion centers (EFC) in sensor networks are generally modeled as unauthorized receivers that passively wiretap communications between sensors and the LFC, have unbounded computational power just like the LFC, and seek to compete against the LFC \cite{mukherjee2015physical}. Due to the optimality of the SPRT, it is natural for the EFC to also employ the SPRT to make its decision.

In order to mitigate the security threats, different approaches have been investigated in recent literature such as stochastic encryption, artificial noise injection, and cooperative jamming, to name a few \cite{kailkhura2015distributed, mukherjee2015physical}.  In particular, as a low-complexity physical-layer security technique, stochastic encryption \cite{aysal2008sensor, soosahabi2012scalable, soosahabi2014optimal} can be employed at the sensors such that every quantized data is transformed according to certain probabilistic rule before transmitted to the LFC. The probabilistic transformation is known only to the LFC, and the EFC is completely ignorant of the existence of stochastic encryptions.
It is worth mentioning that employing stochastic encryptions is an easy way to provide physical layer security for sequential detection with quantized data which does not introduce any communication overhead for the sensors and has minimal processing requirements, rendering it scalable in terms of network size.  In this paper, 
we investigate the sequential detection performance of the EFC in the presence of stochastic encryption and optimize the encryption scheme to  maximize the difference between the expected sample sizes at the EFC and  LFC.

%
%
%

\subsection{Summary of Results}

Since the EFC is unaware of the stochastic encryption, a mismatched SPRT (MSPRT) is employed at the EFC. We characterize the expected sample size and the error probabilities of the MSPRT in terms of the detection thresholds. We show that when the detection error probabilities are set to be the same at the LFC and EFC, every symmetric stochastic encryption leads to the same expected sample size at the LFC and EFC. In addition, the asymptotic analysis on the expected sample size in terms of the vanishing error probabilities is provided, and the stark difference from the asymptotic performance of the SPRT with no model mismatch is revealed. For example, the expected sample size  of the SPRT is  determined by the Kullback-Leibler  (KL) divergences between the distributions under the two hypotheses, while the expected sample size of the MSPRT is unrelated to the KL divergences.  

Next, in the asymptotic regime of small error probabilities, we show that every stochastic encryption degrades the performance of the quantized sequential detection at the LFC by increasing the expected sample size, and the expected sample size required at the EFC is no fewer than that is required at the LFC. Hence, symmetric stochastic encryptions are the least effective ones. Then the optimal stochastic encryption is investigated in the sense of maximizing the difference between the expected sample sizes required at the EFC and LFC.  The optimization problem  is nonconvex. However we show that if the acceptable tolerance of the increase in the expected sample size at the LFC induced by the stochastic encryption is small enough, then the globally optimal stochastic encryption can be analytically obtained. Moreover, the optimal scheme only flips one type of quantized bits (i.e., $1$ or $0$) that has larger probability and keeps the other type unchanged.

\subsection{Related Works}

The decentralized sequential detection in sensor networks using quantized data  has been widely investigated, see \cite{hashemi1989decentralized, veeravalli1993decentralized, veeravalli1999sequential, mei2008asymptotic, cheng2005bandwidth, fellouris2011decentralized, yilmaz2013channel,  li2015multi, li2015decentralized} for instance. However, to the best of our knowledge, stochastic encryption for quantized sequential detection in the presence of eavesdroppers has not been considered.

Stochastic encryptions were originally proposed in \cite{aysal2008sensor} for physical-layer security in the context of  fixed-sample-size estimation problems with quantized data.  For the fixed-sample-size hypothesis testing in sensor networks, the  joint design of the stochastic encryption and the LFC decision rule that minimizes the LFC detection error probabilities subject to a constraint on the EFC error probabilities is studied in \cite{soosahabi2012scalable}. Nonetheless, the design approach  in \cite{soosahabi2012scalable} is ad hoc and results in a suboptimal stochastic encryption. This design approach is made more rigorous in \cite{soosahabi2014optimal}, where the optimal stochastic encryption is obtained with respect to the J-divergence which is adopted as the performance metric for both LFC and EFC. However, 
the approach proposed in \cite{soosahabi2014optimal} cannot be applied to sequential detection.

From the EFC perspective, the stochastic encryption process can be treated as a malicious man-in-the-middle attack \cite{marano2009distributed, vempaty2013distributed, zhang2015Asymptotically, kailkhura2015distributed2}, since the EFC is unaware of the probabilistic transformation of the quantized sensor data.
However,  most  existing works on the man-in-the-middle attacks focus on the fixed-sample-size inference problems, and do not jointly consider the performance at the LFC and EFC. 



The remainder of the paper is organized as follows. The system model and stochastic encryption is introduced in {Section \ref{Section_System_Model}}. 
The performance of the mismatched SPRT employed by the EFC is analyzed in 
{Section \ref{Section_Asymptotic_Characterization}}. In {Section \ref{Section_Optimal_SE}}, the optimal stochastic encryption is pursued. 
Finally, {Section \ref{Section_Conclusion}} provides our conclusions.

\section{System Model and Stochastic Encryption}
\label{Section_System_Model}

In this section, the system model and the stochastic encryption are  introduced. The sequential decision procedures adopted at the LFC and EFC are also specified.

\subsection{Quantized Sequential Detection}

Consider a sensor network consisting of an LFC and $N$ spatially distributed sensors, which aims to test between two hypotheses. Each sensor sequentially makes observations of a particular phenomenon. Let $x_{k}^{(n)}$ denote the $k$-th observation made at the $n$-th sensor. Under each hypothesis, the observations $\{x_{k}^{(n)}\}_{n, k}$ are assumed to be independent and identically distributed (i.i.d.) at each sensor and  are independent across sensors. We use ${f_0^{(n)}}\left( x \right)$ and ${f_1^{(n)}}\left( x \right)$ to denote the probability density functions (pdf) of the observations under the two hypotheses, i.e.,  for all $n \in \{1,2,...,N\}$ and for all  $k \in \{1,2,...\}$
\begin{equation} \label{Hypothesis_testing_problem}
\begin{split}
{{\cal H}_0}: & \; x_{k}^{(n)} \sim {f_0^{(n)}}\left( x \right), \\
{\text{and}} \quad  {{\cal H}_1}: &  \; x_{k}^{(n)} \sim {f_1^{(n)}}\left( x \right).
\end{split}
\end{equation} 

For each observation $x_k^{(n)}$, the $n$-th sensor first forms a one-bit summary message $u_k^{(n)}$ by applying a  quantizer ${\cQ}_n(x) \triangleq \bone_{\{x \in \cD_n\}}$ with the quantization region $\cD_n$, i.e., 
\begin{equation}
u_k^{(n)} \triangleq {\cQ_n} \lp x_k^{(n)} \rp \in \{0,1\},
\end{equation}
and then sends $u_k^{(n)}$ to the LFC.

It is clear that $\{u_k^{(n)}\}$  are independent and identically distributed at each sensor and  are independent across sensors.  Let $p_n$ and $q_n$ denote the probabilities of the event $\{u_k^{(n)} = 1\}$ under the hypotheses $\cH_1$ and $\cH_0$, respectively, i.e.
\begin{equation}
{p_n} \buildrel \Delta \over = \int_{x \in {{\cal D}_n}} {f_1^{(n)}\left( x \right)dx} \text{  and  } {q_n} \buildrel \Delta \over = \int_{x \in {{\cal D}_n}} {f_0^{(n)}\left( x \right)dx}.
\end{equation}
Then the log-likelihood ratio (LLR) of the quantized data $u_k^{(n)}$ can be computed as
\begin{equation} \label{LLR_u_k_n}
l_k^{(n)} = \left\{ \begin{split}
& \ln \frac{{{p_n}}}{{{q_n}}}, \text{  if  } u_k^{(n)}=1, \\
& \ln \frac{{1 - {p_n}}}{{1 - {q_n}}}, \text{ if } u_k^{(n)}=0.
\end{split} \right.
\end{equation}


With regard to the probabilities $\{p_n\}$ and $\{q_n\}$, the following assumption is made throughout the paper.
\begin{assumption} \label{Assumption_pq}
We assume that the local quantizers $\{{\cQ}_n\}_{n=1}^N$ bring about $p_n=p > 0.5$, $q_n=q<0.5$, and $p+q=1$ for all $n \in \{1,2,...,N\}$.
\end{assumption}

It is worth mentioning that \emph{Assumption \ref{Assumption_pq}} is motivated by the classical problem of detecting the mean shift in Gaussian noise \cite{cheng2005bandwidth, sahu2016distributed}.
Specifically, assume the following model and equal priors on both hypotheses 
\begin{equation} \label{problem_example}
\begin{split}
{{\cal H}_0}:  & \; x_{k}^{(n)} = w_{k}^{(n)}, \\
{\text{and}} \quad {{\cal H}_1}:  & \; x_{k}^{(n)} = \theta + w_{k}^{(n)},
\end{split}
\end{equation}
where $\theta$ is a deterministic quantity, and the independent noise $w_{k}^{(n)} \sim \cN(a, b^2)$.
As claimed in \cite{cheng2005bandwidth}, in the sense of minimizing the expected sample size at the LFC, the one-bit optimal threshold quantizer at each sensor is symmetric, i.e., ${\cD_n =  \{{x_k^{(n)}}:  {x_k^{(n)}}\ge a+ \frac{\theta}{2} \} }$,
if a prescribed upper bound on the false alarm and miss probabilities is given. By employing this quantizer at each sensor, it is easy to show that 
all conditions in \emph{Assumption \ref{Assumption_pq}} are satisfied.


We assume that the LFC can reliably receive the quantized data  from the sensors. 
While sequentially receiving the quantized data from the sensors, the LFC implements a sequential decision procedure to test between the hypotheses in (\ref{Hypothesis_testing_problem}). 
Besides the LFC, there exists an EFC in the sensor system which is able to  wiretap the quantized sensor data transmitted from the sensors to the LFC, and the EFC also aims to perform sequential detection between the hypotheses in (\ref{Hypothesis_testing_problem}).
Our goal is to design a strategy at the sensors to transform the quantized data $\{ u_k^{(n)} \}$  so that under the same detection performance constraint, the LFC will reach the decision faster than the EFC since the former is  aware of the transformation but the latter is not. 





\subsection{Stochastic Encryption and the SPRT at the LFC}
\label{SEC_SEandSPRT}


The idea is to stochastically encrypt the quantized data $\{u_k^{(n)}\}$ at each sensor before they are transmitted to the LFC. To be specific, at each sensor, an encrypted version $\tu_k^{(n)}$ of the quantized data $u_k^{(n)}$ is reported to the LFC which follows the encryption rule
\begin{equation} \label{Encryption_Rule}
\lcb \begin{split}
{\mathbbm{P}} \lp \tu_k^{(n)} = 1 \lbar u_k^{(n)} = 0 \rclose \rp =  \psi_0, \\
{\mathbbm{P}} \lp \tu_k^{(n)} = 0 \lbar u_k^{(n)} = 1 \rclose \rp =  \psi_1, 
\end{split} \rclose
\end{equation}
where  $\psi_0$, $\psi_1 \in [0,1]$. 
Hence the quantized data $u_k^{(n)}$ is flipped with probability $\psi_i$ if $u_k^{(n)}=i$ for $i\in\{0,1\}$. Let $\tp$ and $\tq$ denote the probabilities of the event $\{\tu_k^{(n)} = 1\}$ under the hypotheses $\cH_1$ and $\cH_0$, respectively. 
Then we can obtain 
\begin{align} \label{Define_tp}
\tp & = \lp 1 - \psi_0 - \psi_1 \rp p + \psi_0, \\\label{Define_tq}
{\text{   and   }} \tq & = \lp 1 - \psi_0 - \psi_1 \rp q + \psi_0.
\end{align}

We assume that the LFC is aware of the encryption parameters $\psi_0$ and $\psi_1$, while the EFC does not know the existence of the stochastic encryptions.
It is worth mentioning that in order to guarantee this assumption to hold, the encryption parameters $\psi_0$ and $\psi_1$ should be appropriately chosen so that it is hard for the EFC to perceive the existence of the stochastic encryption. More details about the constraints on the encryption parameters $\psi_0$ and $\psi_1$ will be discussed later.

Under \emph{Assumption \ref{Assumption_pq}}, every encrypted bit received at the LFC is independent and follows the same distribution. Hence, for notational simplicity, we use $\tu_t$ to denote the $t$-th encrypted bit received at the LFC henceforth. 
In general, the sequential detection procedure employed by the LFC consists of a stopping rule $\sT_\teL$ and a decision function $\sD_\teL$. The stopping rule  $\sT_\teL$ specifies when the sequential test stops for decision, and 
upon stopping at $\sT_\teL$, 
the decision function $\sD_\teL$ chooses between the two hypotheses.  
The optimal sequential detector ($\sT_\teL$, $\sD_\teL$) minimizes the expected number of sensor data required to reach a decision with probabilities of false alarm and miss upper bounded by $\alpha_\teL^*$ and $\beta_\teL^*$, respectively. 
It is well known that Wald's sequential probability ratio test (SPRT) achieves this optimality \cite{wald1948optimum}. Thus, we assume that the LFC employs the SPRT with the test statistic
\begin{equation} \label{Test_Statistic_LFC}
{{\tilde L}_t} \buildrel \Delta \over = \sum\limits_{s = 1}^t {{{\tilde l}_s}},
\end{equation}
where $\tl_s$ denotes the LLR of the $s$-th received encrypted bit, i.e.,
\begin{equation} \label{LLR_tu_s}
\tl_s = \Ind_{\lcb \tu_s = 1 \rcb} \ln \frac{\tp}{\tq} + \Ind_{\lcb \tu_s =0 \rcb} \ln \frac{1 - \tp}{1 - \tq}.
\end{equation}
The stopping rule and the decision function are given respectively by 
\begin{align} \label{Stopping_Rule_LFC}
\sT_\teL  & \triangleq \inf \lcb t  \lbar \tL_t \notin \lp -A_\teL, B_\teL \rp   \rclose \rcb, \\ \label{Decsion_Function_LFC}
{\text{   and   }}  {\sD_\teL} & \buildrel \Delta \over = \left\{ \begin{array}{l}
1, \text{ if }  \tL_{\sT_\teL} \ge B_\teL, \\
0, \text{ if }  \tL_{\sT_\teL} \le -A_\teL,
\end{array} \right. 
\end{align}
where the thresholds $A_\teL$ and $B_\teL$ are chosen such that ${\tbP_0}\left( {{\sD_\teL} = 1} \right) = {\alpha_\teL^*}$ and ${\tbP_1}\left( {{\sD_\teL} = 0} \right) = {\beta_\teL^*}$. The SPRT given by (\ref{Stopping_Rule_LFC})--(\ref{Decsion_Function_LFC}) is optimal in the sense of minimizing both ${{ \tbE}_0}\left\{ {{\sT_\teL}} \right\}$ and ${{ \tbE}_1}\left\{ {{\sT_\teL}} \right\}$, where $\tbP_i$ and ${\tbE}_i\{\}$ denote the probability measure and the expectation operator under $\cH_i$, respectively.

From (\ref{LLR_tu_s}), 
the LLR 
is bounded from above as per $	|\tl_s| \le \max\lcb \left| \ln \frac{\tp}{\tq} \right|,  \left|\ln \frac{1-\tp}{1-\tq}\right| \rcb$,
which yields the following result.
%

\begin{proposition} \label{Proposition_SPRT}
	As $\alpha_\teL^*$ and $\beta_\teL^*$ tend to $0$ in such a way that $\alpha_\teL^*+\beta_\teL^*<1$, $\alpha_\teL^*\ln {\beta_\teL^*} \to 0$ and $\beta_\teL^*\ln{\alpha_\teL^*} \to 0$, 
	the asymptotic performance of the SPRT employed at the LFC is characterized as
	\begin{align} \label{E_0T_L}
		\tbE_0\{\TL\} & = \underbrace{ \frac{ \alpha_\teL^* \ln \frac{\alpha_\teL^*}{1- \beta_\teL^*} + \lp 1 - \alpha_\teL^* \rp \ln \frac{1-\alpha_\teL^*}{\beta_\teL^*}  }{ \tilde q\ln \frac{{\tilde q}}{{\tilde p}} + \left( {1 - \tilde q} \right)\ln \frac{{1 - \tilde q}}{{1 - \tilde p}} } }_{M_{\teL}^{(0)}} +O(1), \\ \label{E_1T_L}
	{\text{and     }}	\tbE_1\{\TL\} & = \underbrace{ \frac{ \beta_\teL^* \ln \frac{\beta_\teL^*}{1- \alpha_\teL^*} + \lp 1 - \beta_\teL^* \rp \ln \frac{1-\beta_\teL^*}{\alpha_\teL^*}  }{ \tilde p\ln \frac{{\tilde p}}{{\tilde q}} + \left( {1 - \tilde p} \right)\ln \frac{{1 - \tilde p}}{{1 - \tilde q}} } }_{M_{\teL}^{(1)}}+O(1).
	\end{align}
\end{proposition}

The proof of \emph{Proposition \ref{Proposition_SPRT}} is omitted here, since it is similar to the proof of \emph{Theorem 3.1.4} in \cite{tartakovsky2014sequential}. It is worth mentioning that as $\alpha_\teL^*$ and $\beta_\teL^*$ go to $0$, $M_{\teL}^{(0)}$ and $M_{\teL}^{(1)}$ in  (\ref{E_0T_L}) and (\ref{E_1T_L}) respectively  increase to infinity and dominate the $O(1)$ terms,
and therefore, determine the behavior of $\tbE_0\{\TL\}$ and $\tbE_1\{\TL\}$, respectively.

\section{Mismatched SPRT at the EFC}

The EFC also implements the SPRT but based on a mismatched model, since it is unaware of the existence of the stochastic encryption. We refer to such sequential detection procedure as the mismatched SPRT (MSPRT). 

In this section, we first show that symmetric stochastic encryptions are ineffective, since the expected sample sizes at the LFC and EFC are identical when the detection error probabilities are the same at the LFC and EFC.  Then, we obtain the explicit asymptotic characterization of the expected sample size of the MSPRT.

\subsection{Mismatched SPRT and Ineffective Stochastic Encryptions}


From (\ref{LLR_u_k_n}), the mismatched log-likelihood ratio of the $s$-th encrypted bit $\tu_s$, which is based on the unencrypted data model, can be written as
\begin{equation} \label{LLR_EFC}
l_s = \Ind_{\lcb \tu_s = 1 \rcb} \ln \frac{p}{q} + \Ind_{\lcb \tu_s =0 \rcb} \ln \frac{1 - p}{1 - q}.
\end{equation} 

Then, under \emph{Assumption \ref{Assumption_pq}}, the test statistic of the MSPRT employed at the EFC is given by
\begin{equation}  \label{Test_Statistic_EFC_Simplified}
L_t = \sum\limits_{s = 1}^t {{{l}_s}} = \sum_{s=1}^{t}   \eta \lp \Ind_{\lcb \tu_s = 1 \rcb}  - \Ind_{\lcb \tu_s =0 \rcb} \rp,
\end{equation}
where $\eta \triangleq \ln \frac{p}{1-p}$.

Hence, for a given pair of thresholds $(A_\teE, B_\teE)$, the stopping rule $\sT_\teE $ and the decision function $\sD_\teE$ employed at the EFC are given respectively by 
\begin{align} \notag
\sT_\teE   &  \triangleq \inf \lcb t  \lbar L_t \notin \lp -A_\teE, B_\teE \rp   \rclose \rcb \\ \label{TE_simplified}
& = \inf \lcb t   \lbar   \sum_{s=1}^{t}  \lp \Ind_{\lcb \tu_s = 1 \rcb}   -  \Ind_{\lcb \tu_s =0 \rcb} \rp \notin  \lp  -\frac{\AEE}{\eta}, \frac{\BE}{\eta} \rp  \rclose \rcb, \\  \label{Decsion_Function_EFC}
{\text{and    }}{\sD_\teE} & \buildrel \Delta \over = \left\{ \begin{array}{l}
1, \text{ if }  L_{\sT_\teE} \ge B_\teE,\\
0, \text{ if }  L_{\sT_\teE} \le -A_\teE.
\end{array} \right.
\end{align}
Since the test statistic in (\ref{Test_Statistic_EFC_Simplified}) is  a multiple of $\eta$, $\AEE$ and $\BE$ can be chosen to be multiples of $\eta$ so that no overshoot effect occurs. The false alarm and the miss probabilities of the MSPRT can be written as $\alpha_\teE = \tbP_0(\sD_\teE =1)$ and $\beta_\teE = \tbP_1(\sD_\teE=0)$, respectively.



Naturally, under the same error probability constraints,  the LFC attempts to use less sensor data to reach a decision than the EFC by employing the stochastic encryption. 
However, not all stochastic encryptions can achieve this goal. We first provide a theorem regarding a class of ineffective stochastic encryptions.
\begin{theorem} \label{Theorem_Symmetric_Encryption}
	Under \emph{Assumption \ref{Assumption_pq}}, if a symmetric stochastic encryption is employed, i.e., $\psi_0 = \psi_1$, then $\TL = \TE$ as long as 
the detection error probabilities of the LFC and EFC are the same. 
\end{theorem}
\begin{IEEEproof}
	If 
$\psi_0 = \psi_1$,  by (\ref{Define_tp}) and (\ref{Define_tq}), we have
	\begin{equation}
	\tp + \tq = (1 - \psi_0 - \psi_1) (p+q) + 2 \psi_0 = 1.
	\end{equation}
	As a result, the test statistic of the SPRT at the LFC in (\ref{Test_Statistic_LFC}) can be rewritten as
	\begin{equation} \label{Theorem_1_proof_temp1}
	\tilde{L}_t = \sum_{s=1}^{t} \tilde{\eta} \lp \bone_{\{\tu_s = 1\}} -\bone_{\{\tu_s = 0\}}    \rp,
	\end{equation}
	where $\tilde{\eta} \triangleq \ln \frac{\tp}{1 - \tp}$. By comparing (\ref{Test_Statistic_EFC_Simplified}) with (\ref{Theorem_1_proof_temp1}), we can obtain 
	\begin{equation}
	\tilde{L}_t  = \frac{\tilde{\eta}}{\eta} L_t.
	\end{equation}
	Noting that $\AEE$ and $\BE$ are chosen to be multiples of $\eta$,  no overshoot effect occurs in MSPRT. Hence, we can obtain
	\begin{equation}
	\alpha_\teE = \tbP_0 \lp \sD_\teE = 1 \rp = \tbE_0  \lcb  {\bone}_{\lcb L_{\TE} = \BE \rcb} \rcb = \tbE_1  \lcb e^{-\tilde{L}_{\TE}} {\bone}_{\lcb L_{\TE} = \BE \rcb} \rcb = e^{-\frac{\tilde{\eta}}{\eta} \BE} \lp 1 - \beta_\teE \rp,
	\end{equation}
	which implies
	\begin{equation} \label{Theorem_1_proof_temp3}
	{\BE} = \frac{\eta}{\tilde{\eta}} \ln \frac{1-\beta_\teE}{\alpha_\teE}.
	\end{equation}
	On the other hand,  from (\ref{Stopping_Rule_LFC}) and (\ref{Theorem_1_proof_temp1}), the stopping rule $\TL$ can be simplified to
	\begin{equation} \label{Theorem_1_proof_temp2}
	\TL = \inf \lcb t   \lbar   \sum_{s=1}^{t}  \lp \Ind_{\lcb \tu_s = 1 \rcb}   -  \Ind_{\lcb \tu_s =0 \rcb} \rp \notin  \lp  -\frac{\AL}{\tilde{\eta}}, \frac{\BL}{\tilde{\eta}} \rp  \rclose \rcb,
	\end{equation}
	and hence, $\AL$ and $\BL$ can be chosen to be multiples of $\tilde{\eta}$ so that no overshoot effect occurs. From the definition of the false alarm probability, we can obtain
	\begin{equation}
	\alpha_\teL = \tbP_0 \lp \sD_\teL = 1 \rp = \tbE_0  \lcb  {\bone}_{\lcb \tilde{L}_{\TL} = \BL \rcb} \rcb = \tbE_1  \lcb e^{-\tilde{L}_{\TL}} {\bone}_{\lcb \tilde{L}_{\TL} = \BL \rcb} \rcb = e^{- \BL} \lp 1 - \beta_\teL \rp,
	\end{equation}
	which yields
	\begin{equation}  \label{Theorem_1_proof_temp4}
	\BL = \ln \frac{1-\beta_\teL}{\alpha_\teL}.
	\end{equation}
	It is seen from (\ref{Theorem_1_proof_temp3}) and (\ref{Theorem_1_proof_temp4}) that if $\alpha_\teL=\alpha_\teE$ and $\beta_\teL=\beta_\teE$, then
	\begin{equation}
	\frac{\BE}{\eta}  = \frac{\BL}{\tilde{\eta}}.
	\end{equation}
	Similarly, by employing the definition of the miss probability, we can obtain
	\begin{equation}
	\frac{\AEE}{\eta}  = \frac{\AL}{\tilde{\eta}}.
	\end{equation}
	As a result, by comparing (\ref{TE_simplified}) with (\ref{Theorem_1_proof_temp2}), we know that if $\alpha_\teL=\alpha_\teE$ and $\beta_\teL=\beta_\teE$, then
	\begin{equation}
	\TL = \TE
	\end{equation}
	which concludes the proof.
%
\end{IEEEproof}


\subsection{Expected Sample Size and Detection Performance of the Mismatched SPRT}
\label{Section_Asymptotic_Characterization}

In this subsection, we analyze the expected sample size and the detection performance of the MSPRT for a given pair of thresholds $(\AEE,\BE)$.

In order to obtain $\tbE_i\{ \TE \}$ for $i=0,1$, we will make use of a sequence of stopping times $\{{\sT^{(m)}} \}_{m \in \mathbbm{Z}}$ which are defined as
\begin{align}  \label{Define_Tm}
 {\sT^{(m)}}  \triangleq  \inf \! \lcb t   \lbar   m \! + \! \! \sum_{s=1}^{t}  \lp \Ind_{\lcb \tu_s = 1 \rcb}  \!\!-\!\! \Ind_{\lcb \tu_s =0 \rcb}  \rp \! \notin \!  \lp  \! -\frac{\AEE}{\eta}, \!\! \frac{\BE}{\eta} \rp  \rclose \! \rcb.
\end{align}
It is clear that 
\begin{equation} \label{Boundary_Condition_Tm}
\sT^{(m)} = 0, \text{ if } m\le -\frac{\AEE}{\eta} \text{ or } m \ge \frac{\BE}{\eta}.
\end{equation}

According to the definition of $\sT^{(m)}$ in (\ref{Define_Tm}) and the distribution of $\tu_s$, we can obtain that for any $m \in (  -{\AEE}/{\eta}, {\BE}/{\eta} )$,
\begin{align} \label{E_0_Tm_1}
\tbE_0 \lcb \sT^{(m)} \rcb  = \tq  \tbE_0 \lcb \sT^{(m+1)} \rcb  + (1 - \tq) \tbE_0 \lcb \sT^{(m-1)} \rcb +1.
\end{align}
Furthermore, the boundary condition in (\ref{Boundary_Condition_Tm}) implies
\begin{equation} \label{E_0_Tm_2}
\tbE_0 \lcb \sT^{(-\AEE/\eta)} \rcb = \tbE_0 \lcb \sT^{(\BE/\eta)} \rcb = 0.  
\end{equation}
Solving the recursion given by  (\ref{E_0_Tm_1})--(\ref{E_0_Tm_2}), we can obtain that 
if $\tq \ne \frac{1}{2}$, then
\begin{equation}
\tbE_0 \lcb \sT^{(m)} \rcb =\frac{{\left[ {1 - {{\left( {\frac{{1 - \tilde q}}{{\tilde q}}} \right)}^{m + \frac{{{\AEE}}}{\eta }}}} \right]\frac{{\AEE} + {\BE}}{\eta}}}{{\left( {2\tilde q - 1} \right)\left[ {1 - {{\left( {\frac{{1 - \tilde q}}{{\tilde q}}} \right)}^{\frac{{{\AEE}}}{\eta } + \frac{\BE}{\eta}  } }} \right] }} - \frac{{ m+ \frac{\AEE}{\eta}   }}{{\left( {2\tilde q - 1} \right) }},
\end{equation}
and if $\tq = \frac{1}{2}$, then
\begin{equation}
\tbE_0 \lcb \sT^{(m)} \rcb =\left( {m + \frac{{{A_E}}}{\eta }} \right)\left( {\frac{{{B_E}}}{\eta } - m} \right).
\end{equation}
By comparing (\ref{TE_simplified}) and (\ref{Define_Tm}), we can obtain that $\sT^{(0)} = \TE$, and hence,  if $\tq \ne \frac{1}{2}$, 
\begin{equation} \label{E_0_T_E}
\tbE_0 \lcb \TE \rcb = \frac{{\left[ {1 - {{\left( {\frac{{1 - \tilde q}}{{\tilde q}}} \right)}^{\frac{{{\AEE}}}{\eta }}}} \right]\frac{{\AEE} + {\BE}}{\eta}}}{{\left( {2\tilde q - 1} \right)\left[ {1 - {{\left( {\frac{{1 - \tilde q}}{{\tilde q}}} \right)}^{\frac{{{\AEE}}}{\eta } + \frac{\BE}{\eta}  }  }} \right] }} - \frac{{ \frac{\AEE}{\eta}}}{{\left( {2\tilde q - 1} \right) }},
\end{equation}
and if $\tq = \frac{1}{2}$,
\begin{equation}
\tbE_0 \lcb \TE \rcb = \frac{\AEE \BE}{\eta^2}.
\end{equation}

By employing a similar approach,  we obtain that if $\tq \ne \frac{1}{2}$, 
\begin{equation}  \label{E_1_T_E}
\tbE_1 \lcb \TE \rcb = \frac{{\left[ {1 - {{\left( {\frac{{1 - \tilde p}}{{\tilde p}}} \right)}^{\frac{{{\AEE}}}{\eta }}}} \right]\frac{{\AEE} + {\BE}}{\eta}}}{{\left( {2\tilde p - 1} \right)\left[ {1 - {{\left( {\frac{{1 - \tilde p}}{{\tilde p}}} \right)}^{\frac{{{\AEE}}}{\eta } + \frac{\BE}{\eta}  }  }} \right] }} - \frac{{ \frac{\AEE}{\eta}}}{{\left( {2\tilde p - 1} \right) }},
\end{equation}
and if $\tq = \frac{1}{2}$, 
\begin{equation} \label{E_1_T_E_1_2}
\tbE_1 \lcb \TE \rcb = \frac{\AEE \BE}{\eta^2}.
\end{equation}

Next, we consider the detection performance of the MSPRT. Let $\alpha_\teE$  and $\beta_\teE$ denote the false alarm and miss probabilities of the MSPRT at the EFC, respectively, that is,
\begin{align} \label{alpha_E}
\alpha_\teE \triangleq \tbP_0 \lp L_{\TE} \ge \BE   \rp  =  \tbP_0 \lp  \sum_{s=1}^{\TE}  \lp \Ind_{\lcb \tu_s = 1 \rcb}  - \Ind_{\lcb \tu_s =0 \rcb} \rp  \ge \frac{\BE}{\eta}  \rp,
\end{align}
\begin{align} \label{beta_E}
{\text{and    }} \beta_\teE  \triangleq \tbP_1 \lp L_{\TE} \le \AEE   \rp   =  \tbP_1 \lp  \sum_{s=1}^{\TE}  \lp \Ind_{\lcb \tu_s = 1 \rcb}  - \Ind_{\lcb \tu_s =0 \rcb} \rp  \le -\frac{\AEE}{\eta}  \rp.
\end{align}
Based on the sequence of stopping times $\{{\sT^{(m)}} \}_{m \in \mathbbm{Z} }$, we  also define a sequence of probabilities $\{\alpha^{(m)} \}_{m \in \mathbbm{Z} }$ as
\begin{equation} \label{alpha_m}
\alpha^{(m)} \triangleq \tbP_0 \lp m+  \sum_{s=1}^{\sT^{(m)}}  \lp \Ind_{\lcb \tu_s = 1 \rcb} \! - \! \Ind_{\lcb \tu_s =0 \rcb} \rp  \ge \frac{\BE}{\eta}  \rp,
\end{equation}
which yields that $\forall m \in (- {\AEE}/\eta, \BE/\eta  )$
\begin{equation} \label{alpha_m_1}
{\alpha ^{(m)}} = \tilde q {\alpha ^{(m + 1)}} +  \left( {1 - \tilde q} \right){\alpha ^{(m - 1)}}
\end{equation}
by employing the distribution of $\tu_s$.
What is more, the boundary condition in (\ref{Boundary_Condition_Tm}) implies
\begin{equation} \label{alpha_m_2}
{\alpha ^{( - {\AEE}/\eta )}} = 0 \text{ and } {\alpha ^{({\BE}/\eta )}} = 1.
\end{equation}
Solving the recursion in  (\ref{alpha_m_1})--(\ref{alpha_m_2}),
we can obtain
\begin{equation} \label{alpha_m_expression}
{\alpha ^{(m)}} = \left\{ \begin{split}
& \frac{{1 - {{\left( {\frac{{1 - \tilde q}}{{\tilde q}}} \right)}^{ m + \frac{\AEE}{\eta} }  }}}{{1 - {{\left( {\frac{{1 - \tilde q}}{{\tilde q}}} \right)}^{\frac{{{\AEE} + {\BE}}}{\eta }}}}}, \text{ if } \tq \neq \frac{1}{2}, \\
& \frac{\eta }{{{\AEE} + {\BE}}}\lp m + \frac{\AEE}{\eta} \rp,  \text{ if } \tq = \frac{1}{2}.
\end{split} \right.
\end{equation}
From (\ref{alpha_E}) and (\ref{alpha_m}), we know $\alpha_\teE = \alpha^{(0)}$, and therefore, ${\alpha_\teE}$ can be obtained from (\ref{alpha_m_expression}) as
\begin{equation} \label{alpha_E_result}
{\alpha_\teE} = \left\{ \begin{split}
& \frac{{1 - {{\left( {\frac{{1 - \tilde q}}{{\tilde q}}} \right)}^{ \frac{\AEE}{\eta} }  }}}{{1 - {{\left( {\frac{{1 - \tilde q}}{{\tilde q}}} \right)}^{\frac{{{\AEE} + {\BE}}}{\eta }}}}}, \text{ if } \tq \neq \frac{1}{2}, \\
& \frac{\AEE}{{{\AEE} + {\BE}}},  \text{ if } \tq = \frac{1}{2}.
\end{split} \right.
\end{equation}

Similarly, the miss probability of the MSPRT can be derived as
\begin{equation} \label{beta_E_result}
\beta_\teE  = \left\{ \begin{split}
& \frac{{ - {{\left( {\frac{{1 - \tilde p}}{{\tilde p}}} \right)}^{\frac{{{\AEE} + {\BE}}}{\eta }}} + {{\left( {\frac{{1 - \tilde p}}{{\tilde p}}} \right)}^{\frac{{{\AEE}}}{\eta }}}}}{{1 - {{\left( {\frac{{1 - \tilde p}}{{\tilde p}}} \right)}^{\frac{{{\AEE} + {\BE}}}{\eta }}}}}, \text{ if } \tp \neq \frac{1}{2},  \\
& \frac{{{\BE}}}{{{\AEE} + {\BE}}},  \text{ if } \tp = \frac{1}{2}.
\end{split} \right.
\end{equation}

\subsection{Asymptotic Characterization of the Mismatched SPRT}

It is seen from (\ref{E_0_T_E})--(\ref{E_1_T_E_1_2}), (\ref{alpha_E_result}) and (\ref{beta_E_result}) that $\tbE_i \{ \TE \}$, $\alpha_\teE$ and $\beta_\teE$ can all be exactly expressed in terms of the thresholds $\AEE$ and $\BE$. Note that (\ref{alpha_E_result}) and (\ref{beta_E_result}) are transcendental equations, and therefore, do not admit closed-form expressions of $\AEE$ and $\BE$ in terms of $\alpha_\teE$ and $\beta_\teE$. Thus, it is generally difficult to express $\tbE_i \{ \TE \}$ in terms of $\alpha_\teE$ and $\beta_\teE$. In this subsection, we focus on the asymptotic regime where $\alpha_\teE$ and $\beta_\teE$ tend to zero, and derive the asymptotic expressions of  $\tbE_i \{ \TE \}$ in terms of $\alpha_\teE$ and $\beta_\teE$.


Before proceeding, some consideration on the encryption parameters $\psi_0$ and $\psi_1$ is discussed from the perspective of the EFC.  

It is seen from (\ref{alpha_E_result}) and (\ref{beta_E_result}) that by employing different encryption parameters $\psi_0$ and $\psi_1$ which bring about different $\tp$ and $\tq$, the false alarm and miss probabilities at the EFC can be very different. 
In particular, if $\tq =1/2$,  then as $\AEE $ and $\BE$ tend to infinity with  $\AEE =\BE$, 
\begin{equation} \label{argure_tq_temp_1}
\alpha_\teE  = \frac{1}{2} > 0.
\end{equation}
In addition, if $\tq > 1/2$, then by employing (\ref{alpha_E_result}), we can obtain that
\begin{equation}  \label{argure_tq_temp_2}
\alpha_\teE \ge 1  - \lp  \frac{1-\tq}{\tq} \rp > 0, \text{ for any } \AEE \text{ and } \BE.
\end{equation}
Noticing from  (\ref{argure_tq_temp_1}) and (\ref{argure_tq_temp_2}) that if $\tq \ge 1/2$, then the false alarm probability $\alpha_\teE$ at the EFC  cannot be driven to $0$ as $\AEE $ and $\BE$ increase to infinity with  $\AEE =\BE$. On the other hand, it is well known that for the SPRT, the error probabilities decrease to $0$ as the thresholds increase to infinity \cite{tartakovsky2014sequential}.
Thus, if the encryption parameters are chosen such that $\tq \ge 1/2$, then the MSPRT at the EFC does not obey the elementary property of the SPRT that error probabilities decrease to $0$ as the thresholds increase to infinity. In practice, it is possible for the EFC to perceive the encryption, since 
the EFC can choose sufficiently large $\AEE$ and $\BE$ with $\AEE=\BE$ but still observes false alarms.

On the other hand, if $\psi_0$ and $\psi_1$ are chosen such that $\tq < 1/2$, then according to (\ref{alpha_E_result}), the false alarm probability $\alpha_\teE$ at the EFC  can alway be reduced to $0$ by increasing $\AEE $ and $\BE$ to infinity, which agrees with the property of the SPRT.
Hence it is harder for the EFC to perceive the existence of stochastic encryption.
To this end, the system designer should choose the encryption parameters $\psi_0$ and $\psi_1$ to ensure  $\tq < 1/2$ to hinder the EFC from perceiving the existence of stochastic encryption, so that even if the EFC notices that the transmissions from the sensors are stopped before it makes its decision, the EFC would only think this is because it sets the upper bounds on the error probabilities smaller than that at the LFC,
since the EFC is unaware of $\alpha_{\teL}^*$ and $\beta_{\teL}^*$ at the LFC.


Similarly, by analyzing the property of $\beta_\teE$ in (\ref{beta_E_result}), 
we conclude that the encryption parameters $\psi_0$ and $\psi_1$ should ensure $\tp > 1/2$. Therefore, we make the following assumption.
\begin{assumption} \label{Assumption_pq_12}
	In order to hinder the EFC from perceiving the existence of the stochastic encryption,  
the encryption parameters $\psi_0$ and $\psi_1$ are chosen such that
\begin{align} 
\tp & = \lp 1 - \psi_0 - \psi_1 \rp p + \psi_0 > \frac{1}{2}, \\
{\text{and    }} \tq & = \lp 1 - \psi_0 - \psi_1 \rp q + \psi_0 < \frac{1}{2}.
\end{align}
\end{assumption}

%
%

The following theorem characterizes the asymptotic performance of the MSPRT at the EFC when $\alpha_\teE$ and $\beta_\teE$ tend to zero.
\begin{theorem} \label{Theorem_E_TE}
	Under \emph{Assumptions \ref{Assumption_pq}} and \emph{\ref{Assumption_pq_12}},   the following results hold.
	\begin{enumerate}
		\item As $\alpha_\teE$, $\beta_\teE \to 0$, the thresholds $\AEE$, $\BE \to \infty$.
		\item  The expected sample sizes $\tbE_i(\TE)$ at the EFC  under $\cH_0$ and $\cH_1$ can be respectively written in terms of $\alpha_\teE$ and $\beta_\teE$ as
		\begin{align} \label{E_0_TE_theorem}
		& \tbE_0\lcb  \TE \rcb  = \underbrace{  \frac{1}{1- 2\tq } \lsb  (1 - \alpha_\teE) \log_\mu \frac{1}{\beta_\teE} -\alpha_\teE \log_\nu \frac{1}{\alpha_\teE }  \rsb }_{M_{\teE}^{(0)}}+ {o}(1), \\ \label{E_1_TE_theorem}
		& \tbE_1\lcb  \TE \rcb  = \underbrace{ \frac{1}{2\tp -1} \lsb (1 - \beta_\teE) \log_\nu \frac{1}{\alpha_\teE } - \beta_\teE \log_\mu \frac{1}{\beta_\teE} \rsb }_{M_{\teE}^{(1)}} + o(1), 
		\end{align}
		\begin{equation} \label{Define_mu_nu}
		\text{with    } \mu \triangleq  \frac{\tp}{1-\tp} >1 \text{ and } \nu \triangleq  \frac{ 1- \tq}{\tq} >1.
		\end{equation}
		\item For any given $\psi_0$ and $\psi_1$, we have
		\begin{equation} \label{Lemma_Grad_ETE_temp1}
		\frac{\partial M_{\teE}^{(0)} }{{\partial {\beta_\teE}}}< 0 \text{  and } \frac{\partial M_{\teE}^{(1)}}{{\partial {\alpha_\teE}}} < 0,
		\end{equation}
		\begin{equation} \label{Lemma_Grad_ETE_temp2}
		\frac{\partial M_{\teE}^{(0)}}{{\partial {\alpha _\teE}}} < 0 \text{  if  } \alpha_\teE < \frac{1}{e}, 
		\end{equation}
		\begin{equation} \label{Lemma_Grad_ETE_temp3}
		\frac{\partial M_{\teE}^{(1)} }{{\partial {\beta _\teE}}}< 0 \text{  if  } \beta_\teE < \frac{1}{e}.
		\end{equation}
	\end{enumerate} 
\end{theorem}

\begin{IEEEproof}
		We first prove $1)$. Under \emph{Assumptions \ref{Assumption_pq}} and \emph{\ref{Assumption_pq_12}}, by employing (\ref{alpha_E_result}), (\ref{beta_E_result}) and (\ref{Define_mu_nu}), the expressions of $\alpha_\teE$ and $\beta_\teE$ can be simplified to 
		\begin{equation} \label{alpha_E_nu}
		\alpha_\teE = \frac{{ {{\nu}^{ \frac{\AEE}{\eta} }  }} -1 }{{{{\nu}^{\frac{{{\AEE} + {\BE}}}{\eta }}}} -1} = \frac{1 - \nu^{-\frac{\AEE}{\eta}} }{\nu^{\frac{\BE}{\eta}} - \nu^{-\frac{\AEE}{\eta}}},
		\end{equation}
		\begin{equation} \label{beta_E_mu}
		\text{and    }\beta_\teE = \frac{{ {{\mu}^{ \frac{\BE}{\eta} }  }} -1 }{{{{\mu}^{\frac{{{\AEE} + {\BE}}}{\eta }}}} -1} = \frac{1 - \mu^{-\frac{\BE}{\eta}} }{\mu^{\frac{\AEE}{\eta}} - \mu^{-\frac{\BE}{\eta}}}.
		\end{equation}
		Since $\AEE / \eta \ge 1$ and $\BE /\eta \ge 1$, $\alpha_\teE$ and $\beta_\teE$ can be bounded from below as per
		\begin{equation}
		\alpha_\teE > \frac{1 - \nu^{-1}}{\nu^{\frac{B_\teE}{\eta}}} \text{ and } \beta_\teE > \frac{1 - \mu^{-1}}{\mu^{\frac{A_\teE}{\eta}}},
		\end{equation}
		which implies that as $\alpha_\teE \to 0$, $\BE \to \infty$,
		and similarly, $\AEE \to \infty$ as $\beta_\teE \to 0$.
		
		Next, we prove $2)$. From (\ref{alpha_E_nu}), we can obtain
		\begin{equation} \label{alpha_E_temp}
		\alpha_\teE \nu^{\frac{\BE}{\eta}} -1 = -(1 - \alpha_\teE)\nu^{-\frac{\AEE}{\eta}}.
		\end{equation}
		Note that $\AEE \to \infty$, as  $\beta_\teE \to 0$, and hence from (\ref{alpha_E_temp}), we know 
		\begin{equation}
		\alpha_\teE \nu^{\frac{\BE}{\eta}} -1 = o(1), \text{ as } \beta_\teE \to 0,
		\end{equation}
		which implies
		\begin{equation} \label{B_E_alpha}
		\frac{\BE}{\eta} = \log_\nu \frac{1}{\alpha_\teE} + o(1), \text{ as }  \beta_\teE \to 0.
		\end{equation}
		
		Similarly, (\ref{beta_E_mu}) yields that 
		\begin{equation} \label{A_E_beta}
		\frac{\AEE}{\eta} = \log_\mu \frac{1}{\beta_\teE} + o(1), \text{ as } \alpha_\teE \to 0.
		\end{equation}
		
		By employing (\ref{E_0_T_E}), (\ref{Define_mu_nu}), (\ref{alpha_E_nu}), (\ref{B_E_alpha}) and (\ref{A_E_beta}), we can obtain that as $\alpha_\teE$, $\beta_\teE \to 0$
		\begin{align} \notag
		\tbE_0 \{\TE \}  & = \frac{1}{1 - 2\tq } \lsb  (1 - \alpha_\teE) \frac{\AEE}{\eta} -\alpha_\teE \frac{\BE}{\eta}  \rsb \\ \notag
		& =  \frac{1}{1 - 2\tq } \lsb   (1 - \alpha_\teE) \log_\mu \frac{1}{\beta_\teE} - \alpha_\teE \log_\nu \frac{1}{\alpha_\teE }\rsb + {o}(1).
		\end{align}
		
		Similarly, we can show that as $\alpha_\teE$, $\beta_\teE \to  0$,
		\begin{equation} \notag
		\tbE_1\lcb  \TE \rcb = \frac{1}{2\tp -1} \lsb (1 - \beta_\teE) \log_\nu \frac{1}{\alpha_\teE } - \beta_\teE \log_\mu \frac{1}{\beta_\teE} \rsb + o(1).
		\end{equation}
		
		At last, we prove $3)$. By employing the definition of $M_{\teE}^{(0)}$ in  (\ref{E_0_TE_theorem}), we can obtain
		\begin{equation} \label{M_E_0_temp1}
		\frac{\partial M_{\teE}^{(0)} }{{\partial {\alpha _\teE}}}   =  \frac{1}{1-2\tq}\frac{1}{\ln \nu} \lp 1  -   \log_{\mu}\nu \ln \frac{1}{\beta_\teE}    -   \ln \frac{1}{\alpha_\teE} \rp.
		\end{equation}
		Under \emph{Assumption \ref{Assumption_pq_12}}, we know that $\tq < \frac{1}{2}$, and moreover, from (\ref{Define_mu_nu}), we know $\nu >1$. Hence, $\frac{\partial M_{\teE}^{(0)}}{{\partial {\alpha _\teE}}} <0$ if and only if 
		\begin{equation}
		1 - \log_{\mu}\nu \ln \frac{1}{\beta_\teE}  -  \ln \frac{1}{\alpha_\teE} <0,
		\end{equation}
		which is equivalent to 
		\begin{equation} \label{proof_Lemma_Grad_ETE_temp1}
		\alpha_\teE < \frac{1}{e}\lp \frac{1}{\beta_\teE} \rp^{\log_\mu \nu}.
		\end{equation}
		Noting that $\beta_\teE \in [0,1]$ and $\mu,\nu >1$, we know that if $\alpha_\teE < \frac{1}{e}$, then (\ref{proof_Lemma_Grad_ETE_temp1}) is satisfied, and hence, $\frac{\partial M_{\teE}^{(0)}}{{\partial {\alpha _\teE}}} <0$. 
		
		On the other hand, from the definition of $M_{\teE}^{(0)}$ in  (\ref{E_0_TE_theorem}), we can obtain
		\begin{equation} \label{M_E_0_temp2}
		\frac{\partial M_{\teE}^{(0)} }{{\partial {\beta _\teE}}} = -\frac{1}{\lp 1-2\tq \rp \ln \mu} \frac{1 -\alpha_\teE}{\beta_\teE}<0.
		\end{equation}
		
		Similarly, it can be shown that for any given $\psi_0$ and $\psi_1$, $\frac{\partial M_{\teE}^{(1)} }{{\partial {\beta _\teE}}}< 0$ if $\beta_\teE < \frac{1}{e}$ and $\frac{\partial M_{\teE}^{(1)}}{{\partial {\alpha_\teE}}} < 0$.
\end{IEEEproof}

It is seen from the definitions in (\ref{E_0_TE_theorem}) and (\ref{E_1_TE_theorem}) that $M_{\teE}^{(0)}$, $M_{\teE}^{(1)} \to \infty$ as $\alpha_\teE$, $\beta_\teE \to 0$, and hence, dominate the $o(1)$ terms 
and determine the behavior of $\tbE_0\{ \TE \}$ and $\tbE_1\{ \TE \}$, respectively. 


By comparing $M_{\teE}^{(i)}$ in (\ref{E_0_TE_theorem}) and (\ref{E_1_TE_theorem}) with $M_\teL^{(i)}$ in (\ref{E_0T_L}) and (\ref{E_1T_L}), it is clear that 
the dominant term $M_\teE^{(i)}$ in $\tbE_i \{ \TE \}$ couples with the encryption parameters $\psi_0$ and $\psi_1$ in a more complicated way when compared to the dominant term $M_\teL^{(i)}$ in $\tbE_i \{ \TL \}$. In particular, 
$M_\teL^{(i)}$ is  determined by the KL divergences between the distributions under $\cH_0$ and $\cH_1$, while 
$M_\teL^{(i)}$ is unrelated to the KL divergences.



Note that from \emph{Theorem \ref{Theorem_E_TE}}, if $\alpha_\teE < \frac{1}{e}$ and $\beta_\teE < \frac{1}{e}$, then $M_{\teE}^{(i)}$ monotonically decreases as $\alpha_\teE$ or $\beta_\teE$ increases.  On the other hand, by employing the definitions of $M_{\teL}^{(0)}$ and $M_{\teL}^{(1)}$ in (\ref{E_0T_L}) and (\ref{E_1T_L}), after some algebra, we can obtain that if $\alpha_{\teL}^* + \beta_\teL^*<1$, then
\begin{align}
\frac{{\partial M_{\teL}^{(0)} }}{{\partial \alpha_{\teL}^* }} & \propto   \frac{{\partial M_{\teL}^{(1)} }}{{\partial \beta_\teL^* }}  \propto  - \ln \left( {1  +  \frac{{1 - \alpha_{\teL}^*  - \beta_\teL^* }}{{ \alpha_{\teL}^* \beta_\teL^* }}} \right) < 0, \\
\frac{{\partial M_{\teL}^{(0)} }}{{\partial \beta_\teL^* }} &  \propto  \frac{{\partial M_{\teL}^{(1)} }}{{\partial \alpha_{\teL}^* }}  \propto  - \left( {1 - \alpha_{\teL}^*  - \beta_\teL^* } \right) < 0.
\end{align}
which implies that for each $i$, $M_{\teL}^{(i)}$ also monotonically decreases as $\alpha_\teL^*$ or $\beta_\teL^*$ increases. 


\begin{figure}[htb]
	\centerline{
		\includegraphics[width=0.86\textwidth]{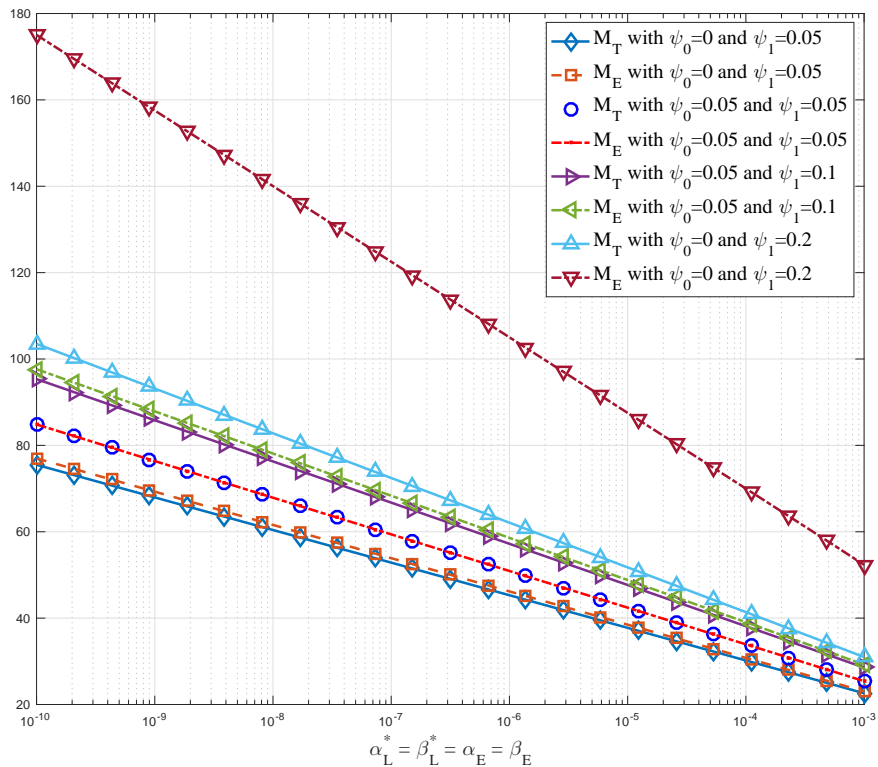}
	}
	\caption{$M_\teL$ and $M_\teE$ under different ($\psi_0$, $\psi_1$).		
	}	
	\label{Fig_MT_ME}	
\end{figure}

To compare the asymptotic performance of the SPRT with that of the MSPRT numerically,
Fig. \ref{Fig_MT_ME} depicts the values of $M_\teL \triangleq \pi_0 M_{\teL}^{(0)} + \pi_1 M_{\teL}^{(1)} $ and $M_\teE \triangleq \pi_0 M_{\teE}^{(0)} + \pi_1 M_{\teE}^{(1)}$ under different stochastic encryptions when $\alpha_\teL^*=\beta_\teL^*=\alpha_\teE=\beta_\teE$ grow from $10^{-10}$ to $10^{-3}$.  We set $p=0.7$, $q=0.3$,  and the priors of $\cH_0$ and $\cH_1$ as  $\pi_0 = \pi_1 =0.5$.
It is seen from Fig. \ref{Fig_MT_ME} that $M_\teL$ and $M_\teE$ both decrease as $\alpha_\teL^*=\beta_\teL^*=\alpha_\teE=\beta_\teE$ increase, and the difference between $M_\teL$ and $M_\teE$ vary considerably for different  stochastic encryption parameters. For example, when $\psi_0=\psi_1=0.05$, $M_\teL=M_\teE$ which agrees with the results in \emph{Theorem \ref{Theorem_Symmetric_Encryption}}, while there is a big gap between $M_\teL$ and $M_\teE$  when $\psi_0=0$ and $\psi_1=0.2$. Furthermore, Fig. \ref{Fig_MT_ME} illustrates that under different stochastic encryptions, the difference between the slope of $M_\teE$ and that of $M_\teL$ can be significantly different. These observations motivate us to pursue the optimal stochastic encryption.

\section{Optimal Stochastic Encryption}
\label{Section_Optimal_SE}

In this section, we consider the optimization of the encryption parameters $\psi_0$ and $\psi_1$ with the goal of maximizing the difference between the expected sample sizes at the EFC and LFC with probabilities of false alarm and miss upper bounded by prescribed values. 
For a fair comparison between the expected sample sizes at the LFC and EFC, we set the upper bounds on the error probabilities to be identical at the LFC and EFC, i.e., $\alpha_\teL \le \alpha^*$, $\alpha_\teE \le \alpha^*$, $\beta_\teL \le \beta^*$ and $\beta_\teE \le \beta^*$.
Moreover, denote $\bPsi \triangleq [\psi_0, \psi_1]$.


\subsection{Optimization Formulation}

To take into account the  increase in the expected sample sizes at the LFC   induced by the stochastic encryption, we impose the following constraints
\begin{equation} \label{Define_lambda}
{\lambda}_i \lp \bPsi\rp \triangleq \frac{\tbE_i\{ \TL \} -\bbE_i\{ \TL \}}{\bbE_i\{ \TL \}} \le \kappa_i,  i=0,1,
\end{equation}
where  $\kappa_i$ is a nonnegative constant which represents the upper bound on the acceptable tolerance of the increase in the expected sample sizes at the LFC induced by the stochastic encryption. The term $\bbE_i\{ \TL \}$ in (\ref{Define_lambda}) corresponds to the case of 
no stochastic encryption, which can be obtained from (\ref{E_0T_L}) and (\ref{E_1T_L}) by replacing $\tp$ and $\tq$ by $p$ and $q$, respectively.



Under \emph{Assumption \ref{Assumption_pq_12}}, 
the optimization of the stochastic encryption parameter $\bPsi$ can be cast as the following maximin problem

\begin{subequations} \label{Optimal_SE_Problem}
	\begin{align} \label{Optimal_SE_Problem_Objective}
	\mathop {\max }\limits_{\bPsi ,{\alpha _\teL},{\beta _\teL}} \; & \mathop {\min }\limits_{{\alpha _\teE},{\beta _\teE}} \; \sum\limits_{i = 0}^1 {{\pi _i}\left( {{\tbE_i}\left\{ {{\TE}} \right\} - {\tbE_i}\left\{ {{\TL}} \right\}} \right)}  \\ \label{Optimal_SE_Problem_Constraint_Delay}
	\text{s. t.} \quad  & {\lambda}_i \lp \bPsi\rp \le \kappa_i,  \; \forall i=0,1, \\ \label{Optimal_SE_Problem_Constraint_tp}
	& \lp 1 - \psi_0 - \psi_1 \rp p + \psi_0  > \frac{1}{2}, \\  \label{Optimal_SE_Problem_Constraint_tq}
	&  \lp 1 - \psi_0 - \psi_1 \rp q + \psi_0  < \frac{1}{2},  \\  \label{Optimal_SE_Problem_Constraint_alpha_beta_LFC}
	& \alpha_\teL \le {\alpha ^*}, \; \beta_\teL \le {\beta ^*}, \\ \label{Optimal_SE_Problem_Constraint_alpha_beta_EFC}
	& \alpha_\teE \le {\alpha ^*}, \; \beta_\teE \le {\beta ^*}.
	\end{align}
\end{subequations}

Since the LFC employs the SPRT, it is clear that the constraint in (\ref{Optimal_SE_Problem_Constraint_alpha_beta_LFC}) is active for the optimal solution, that is,
\begin{equation}
\alpha_\teL = {\alpha ^*}  \text{ and }  \beta_\teL = {\beta ^*}.
\end{equation}
However, for the MSPRT, the expected sample sizes $\tbE_i \{ \TE \}$ may not be minimized when $\alpha_\teE = \alpha^*$ and $\beta_\teE = \beta^*$ in general.  To this end, as illustrated in the maximin problem in (\ref{Optimal_SE_Problem}), we consider the best performance of the EFC in terms of the expected sample size when its detection performance satisfies the constraints, that is, $\alpha_\teE \le \alpha^*$ and $\beta_\teE \le \beta^*$.

\subsection{Optimization Problem in the Asymptotic Regime}

In general, the optimization problem in (\ref{Optimal_SE_Problem}) is not tractable, since the closed-form expression for the objective function in (\ref{Optimal_SE_Problem_Objective}) does not generally exist.  In the following, we consider the objective function in (\ref{Optimal_SE_Problem_Objective}) in the asymptotic regime where $\alpha^*$ and $\beta^*$ are sufficiently small. From  (\ref{E_0T_L}), (\ref{E_1T_L}), (\ref{E_0_TE_theorem}) and (\ref{E_1_TE_theorem}), 
by keeping the leading-order terms and ignoring the lower-order terms,  $\tbE_i\{ \TL \}$ and $\tbE_i \{ \TE \}$ can be approximated in the asymptotic regime as
\begin{equation} \label{E_0_T_L_approx}
\tbE_0\{\TL\} \approx \frac{  - \ln {\beta^*}   }{ \tilde q\ln \frac{{\tilde q}}{{\tilde p}} + \left( {1 - \tilde q} \right)\ln \frac{{1 - \tilde q}}{{1 - \tilde p}} } \triangleq {\hTLZ},
\end{equation}
\begin{equation} \label{E_1_T_L_approx}
\tbE_1\{\TL\} \approx \frac{ - \ln {\alpha^*} }{ \tilde p\ln \frac{{\tilde p}}{{\tilde q}} + \left( {1 - \tilde p} \right)\ln \frac{{1 - \tilde p}}{{1 - \tilde q}} } \triangleq {\hTLO},
\end{equation}
\begin{equation} \label{E_0_T_E_approx}
\tbE_0\lcb  \TE \rcb \approx \frac{1}{1 - 2\tq} \log_\mu \frac{1}{\beta_\teE} \triangleq {\hTEZ},
\end{equation}
\begin{equation} \label{E_1_T_E_approx}
\text{and  } \tbE_1\lcb  \TE \rcb \approx \frac{1}{2\tp -1}\log_\nu \frac{1}{\alpha_\teE } \triangleq {\hTEO}.
\end{equation}
It is worth mentioning that the asymptotic approximations employed in (\ref{E_0_T_L_approx})--(\ref{E_1_T_E_approx}) are similar to that suggested by Wald in \cite{wald1947sequential} which is known as \emph{Wald's approximation}, and are commonly utilized in recent literature, see \cite{cohen2015asymptotically, bai2015stochastic} for instance. 

It is seen from (\ref{E_0_T_E_approx}) and (\ref{E_1_T_E_approx}) that $\hTE^{(i)}(\bPsi)$ is nonincreasing functions of $\alpha_\teE$ and $\beta_\teE$, and therefore, the optimization problem in (\ref{Optimal_SE_Problem}) can be reduced to 

\begin{subequations} \label{Optimal_SE_Problem_no_inf}
	\begin{align} \label{Optimal_SE_Problem_Objective_no_inf}
	\mathop {\max }\limits_{\bPsi} \quad  &  \pi_0 \lsb \hTE^{(0)}(\bPsi) - \hTL^{(0)}(\bPsi) \rsb +  \pi_1 \lsb \hTE^{(1)}(\bPsi) - \hTL^{(1)}(\bPsi) \rsb \\ \label{Optimal_SE_Problem_Constraint_Delay_no_inf}
	\text{s. t.} \quad  & \hlambda_i \lp \bPsi\rp \le \kappa_i,  \; \forall i=0,1, \\ \label{Optimal_SE_Problem_Constraint_tp_no_inf}
	& \lp 1 - \psi_0 - \psi_1 \rp p + \psi_0  > \frac{1}{2}, \\  \label{Optimal_SE_Problem_Constraint_tq_no_inf}
	&  \lp 1 - \psi_0 - \psi_1 \rp q + \psi_0  < \frac{1}{2},  \\  \label{Optimal_SE_Problem_Constraint_alpha_beta_no_inf}
	& \alpha_\teL = \alpha_\teE = {\alpha ^*}, \; \beta_\teL = \beta_\teE =  {\beta ^*},
	\end{align}
\end{subequations}
where 
\begin{equation} \label{Define_hlambda}
\hlambda_i(\bPsi) \triangleq \frac{\hTLi - \hTL^{(i)}([0,0])}{\hTL^{(i)}([0,0])}, i=0,1,
\end{equation}  
is the corresponding asymptotic approximation of ${\lambda}_i(\bPsi)$ in (\ref{Define_lambda}).

By plugging (\ref{E_0_T_L_approx})--(\ref{E_1_T_E_approx}) into (\ref{Optimal_SE_Problem_no_inf}), we attain an optimization problem where every term has an  analytic expression. However, the optimization problem in (\ref{Optimal_SE_Problem_no_inf}) is generally nonconvex. In particular, both the objective function in (\ref{Optimal_SE_Problem_Objective_no_inf}) and the feasible region specified by (\ref{Optimal_SE_Problem_Constraint_Delay_no_inf})--(\ref{Optimal_SE_Problem_Constraint_tq_no_inf}) are generally nonconvex. Thus, it is generally intractable to find the globally optimal solution.

\subsection{Optimal Solution under Small $\kappa_i$}

In this subsection, we will show that 
for small $\kappa_i$ in (\ref{Optimal_SE_Problem_Constraint_Delay_no_inf}), the globally optimal solution to (\ref{Optimal_SE_Problem_no_inf}) can be analytically obtained.

We first look into the constraint in (\ref{Optimal_SE_Problem_Constraint_Delay_no_inf}). By employing (\ref{E_0_T_L_approx}) and (\ref{E_1_T_L_approx}), $\hlambda_i(\bPsi)$ can be rewritten as
\begin{equation} \label{lambda_simplified}
\hlambda_0\lp  \bPsi \rp = \frac{H(q,p)}{H(\tq,\tp)} -1 \text{ and } \hlambda_1\lp  \bPsi \rp = \frac{H(p,q)}{H(\tp,\tq)} -1,
\end{equation}
\begin{equation} \label{Define_H}
\text{with  } H(x,y) \triangleq x \ln \frac{x}{y} + (1-x) \ln \frac{1-x}{1-y} > 0, \text{ if } x \ne y.
\end{equation}

The following Lemma provides some insights into the constraint in (\ref{Optimal_SE_Problem_Constraint_Delay_no_inf}). The proof is given in Appendix \ref{Proof_Lemma_Grad_lambda}. 
\begin{lemma} \label{Lemma_Grad_lambda}
	Under \emph{Assumptions \ref{Assumption_pq}} and \emph{\ref{Assumption_pq_12}}, we have the following results.
	\begin{enumerate}
		\item For any given $\bPsi$, 
		\begin{equation} \label{Lemma_Grad_lambda_temp1}
		{\nabla _\bPsi }\hlambda_i \left( \bPsi  \right) \succ \boldsymbol{0}, \;  i=0,1.
		\end{equation}
		\item There exist two constants $\zeta_\hlambda$ and $c_{\hlambda}$ such that if 
		\begin{equation} \label{Lemma_zeta_lambda}
		\kappa \triangleq \max\{\kappa_0, \kappa_1\} <  \zeta_\hlambda
		\end{equation}
		then
		\begin{align} \label{Define_psi_lambda}
		\psi_j^{(\hlambda)} & \triangleq  \sup \lcb \psi_j | \exists \psi_{1-j}, \text{ s.t. } \hlambda_i \lp \bPsi\rp \le \kappa_i,  \;  i=0,1 \rcb \\ \label{Lemma_Grad_H_sup_simplified_1}
		&  =  \sup \lcb \psi_j |  \hlambda_i \lp \bPsi \rp \le \kappa_i \text{ with }  \psi_{1-j}=0, \;  i=0,1 \rcb \\ \label{Lemma_Grad_H_sup_simplified_2}
		&  =  \min \lcb \psi_{j, \hlambda_0}, \psi_{j, \hlambda_1}\rcb \\ \label{Lemma_Grad_H_sup_simplified_3}
		&  <  c_\hlambda \kappa,
		\end{align}
		where 
		$\psi_{j, \hlambda_i}$  is the  solution to $\hlambda_i \lp \bPsi \rp = \kappa_i$ given $\psi_{1-j}=0$.
	\end{enumerate}
%
\end{lemma}


The constants $\zeta_{\hlambda}$ and $c_{{\hlambda}}$ in \emph{Lemma \ref{Lemma_Grad_lambda}} are defined in (\ref{proof_Lemma_zeta_lambda}).
Using the definition of $\psi_j^{({\hlambda})}$ in (\ref{Define_psi_lambda}), 
denote the following two points in the $\psi_0$-$\psi_1$ plane, 
\begin{equation} \label{Define_Psi_01_lambda}
\bPsi_0^{({\hlambda})} \triangleq \lsb\psi_0^{({\hlambda})}, 0 \rsb \text{  and  } \bPsi_1^{({\hlambda})} \triangleq  \lsb 0, \psi_1^{({\hlambda})} \rsb.
\end{equation}

Noticing from (\ref{Define_hlambda}), it is clear that $\hlambda_i([0,0])=0, i=0,1$. Moreover, as  demonstrated by 1) in \emph{Lemma \ref{Lemma_Grad_lambda}}, ${\nabla _\bPsi }\hlambda_i \left( \bPsi  \right) \succ \boldsymbol{0}$ which implies that  if $\bPsi \ne {\bf{0}}$, then $\hlambda_i(\bPsi)>0, i=0,1$, and hence, $\hTLi > \hTL^{(i)}([0,0]), i=0, 1$. Therefore, every stochastic encryption degrades the performance of the SPRT at the LFC by increasing the expected sample size. Furthermore, as 2) in \emph{Lemma \ref{Lemma_Grad_lambda}} illustrates, the points $\bPsi_0^{({\hlambda})}$ and $\bPsi_1^{({\hlambda})}$ respectively attain the largest possible values of $\psi_0$ and $\psi_1$ in the set specified by $\cap_{i = 0}^1 \{ {\bPsi | {{{\hlambda} _i}( \bPsi  ) \le {\kappa_i}} } \}$.
Moreover, these two largest values $\psi_0^{({\hlambda})}$ and $\psi_1^{({\hlambda})}$ are bounded from above and can be controlled by $\kappa$.
It is worth mentioning that 
the set specified by ${\hlambda}_i \left( \bPsi  \right)  \le \kappa_i$ is just the region enclosed by $[0, \psi_0^{({\hlambda})}] \times \{0\}$,  $\{0\} \times [0, \psi_1^{({\hlambda})} ]$ and the contour of ${\hlambda}_i \left( \bPsi  \right)  = \kappa_i$.

Next we consider the constraints (\ref{Optimal_SE_Problem_Constraint_tp_no_inf}) and (\ref{Optimal_SE_Problem_Constraint_tq_no_inf}). Let $[0, \psi_1^{(\tp)} ]$ and $[\psi_0^{(\tq)}, 0 ]$ respectively denote the point where the line $\lp 1 - \psi_0 - \psi_1 \rp p + \psi_0  = 1/2$ intersects the $\psi_1$-axis and the point where the line $\lp 1 - \psi_0 - \psi_1 \rp q + \psi_0  = 1/2$ intersects the $\psi_0$-axis.
It can be shown that
\begin{equation} \label{psi_tp_tq}
\psi_1^{(\tp)} = 1 - \frac{1}{2p} \text{  and } \psi_0^{(\tq)} = \frac{1-2q}{2(1-q)}.
\end{equation}
According to the constraints in (\ref{Optimal_SE_Problem_Constraint_tp_no_inf}) and (\ref{Optimal_SE_Problem_Constraint_tq_no_inf}), we know that in the $\psi_0$-$\psi_1$ plane, the closed interval $\{0\} \times [0, \psi_1^{(\tp)} ]$ 
and the closed interval $[0, \psi_0^{(\tq)}] \times \{0\}$ 
are  both contained in the set specified by (\ref{Optimal_SE_Problem_Constraint_tp_no_inf}) and (\ref{Optimal_SE_Problem_Constraint_tq_no_inf}). Therefore, 
if \begin{equation} \label{kappa_smaller_zeta_pq}
\kappa < \zeta_{\tp,\tq} \triangleq \min \lcb   \frac{\psi_0^{(\tq)}}{c_{{\hlambda}}},  \frac{\psi_1^{(\tp)}}{c_{{\hlambda}}} \rcb,
\end{equation} 
then $\psi_0^{({\hlambda})} < \psi_0^{(\tq)}$ and $\psi_1^{({\hlambda})} < \psi_1^{(\tq)}$, and hence\footnote{If $j=0$, then the vector ${[{\psi _{1 - j}},{\psi _j}]}$ in (\ref{psi_j_lambda_sup}) needs to be replaced by ${[{\psi _{j}},{\psi _{1-j}}]}$.}
\begin{equation} \label{psi_j_lambda_sup}
\sup \left\{ {{\psi _j}\left| {\exists {\psi _{1 - j}},{\text{ s.t. }}\bPsi  = {{[{\psi _{1 - j}},{\psi _j}]}} \in {\mathcal{E}}} \right.} \right\} = \psi _j^{({\hlambda} )},
\end{equation}
where $\mathcal{E}$ denotes the feasible set specified by all the constraints in the optimization problem in (\ref{Optimal_SE_Problem_no_inf}). From (\ref{psi_j_lambda_sup}), we know that
\begin{equation} \label{Define_bar_E}
{\cE} \subseteq {\bar{\cE}} \triangleq \lcb \bPsi : \psi_0 \in [0, \psi_0^{(\hlambda)}], \psi_1 \in [0, \psi_1^{(\hlambda)}]  \rcb.
\end{equation}


With regard to the behavior of the objective function in (\ref{Optimal_SE_Problem_Objective_no_inf}), we have the following lemma. The proof is given in Appendix \ref{Proof_Lemma_Grad_Objective}.
\begin{lemma} \label{Lemma_Grad_Objective}
Under \emph{Assumptions \ref{Assumption_pq}} and \emph{\ref{Assumption_pq_12}}, we have the following results.
\begin{enumerate}
	\item  $\hTEi - \hTLi \ge 0$, $ i=0,1$,
	with  equality 
	if and only if a symmetric encryption is employed, i.e., $\psi_0 = \psi_1$. 
	\item There exists a constant $\zo$ such that if 
	\begin{equation}
	\kappa < \zo,
	\end{equation}
	then  in the region  ${\bar{\cE}}\cap \{ \bPsi: \psi_1 > \psi_0 \ge 0 \}$,  we have that 
	\begin{equation} \label{Grad_Objective_psi1_bigger_psi0}
	\frac{\partial \lsb \hTEi- \hTLi \rsb }{\partial \psi_0} < 0 \text{  and  } \frac{\partial \lsb \hTEi- \hTLi \rsb }{\partial \psi_1} > 0,   \; i=0,1,
	\end{equation}
	while in the region  ${\bar{\cE}} \cap \{ \bPsi: 0 \le \psi_1 < \psi_0\}$, we have that 
	\begin{equation} \label{Grad_Objective_psi1_smaller_psi0}
	\frac{\partial \lsb \hTEi- \hTLi \rsb }{\partial \psi_0} > 0 \text{  and  } \frac{\partial \lsb \hTEi- \hTLi \rsb }{\partial \psi_1} < 0, \;  i=0,1.
	\end{equation}
\end{enumerate}
\end{lemma}


As illustrated by 1) in \emph{Lemma \ref{Lemma_Grad_Objective}}, the expected sample size at the EFC is no fewer than that at the LFC in the asymptotic regime where $\alpha^*, \beta^* \to 0$. From \emph{Theorem \ref{Theorem_Symmetric_Encryption}} and \emph{Lemma \ref{Lemma_Grad_Objective}}, we have the following corollary regarding the symmetric stochastic encryptions.
\begin{corollary} \label{Corollary_symmetric_encryption}
	Under \emph{Assumptions \ref{Assumption_pq}} and \emph{\ref{Assumption_pq_12}}, the symmetric stochastic encryptions are the least favorable in the asymptotic regime where $\alpha^*$, $\beta^* \to 0$, since they are the only class of stochastic encryptions which cannot help the LFC outperform the EFC in terms of the expected sample sizes.
\end{corollary}


Finally, in the following theorem, for small $\kappa$, we give the optimal stochastic encryption in the sense of maximizing the difference between the expected sample sizes at the EFC and LFC.
\begin{theorem} \label{Theorem_Optimal_Solution}
	If the following conditions 
	\begin{enumerate} 
		\item[(C1)] $\psi_0^{({\hlambda})} \le \frac{1-2q}{2(1-q)}$ and $\psi_1^{({\hlambda})} \le 1- \frac{1}{2p}$
		\item[(C2)] for $i=0,1$, \\
		$\frac{\partial \lsb \hTEi- \hTLi \rsb }{\partial \psi_0} < 0 \text{  and  } \frac{\partial \lsb \hTEi- \hTLi \rsb }{\partial \psi_1} > 0 $  in the region ${\cE}\cap \{ \bPsi: \psi_1 > \psi_0 \ge 0 \}$ \\
		$\frac{\partial \lsb \hTEi- \hTLi \rsb }{\partial \psi_0} > 0 \text{  and  } \frac{\partial \lsb \hTEi- \hTLi \rsb }{\partial \psi_1} < 0 $ in the region  ${\cE} \cap \{ \bPsi: \psi_0 > \psi_1 \ge 0 \}$
	\end{enumerate}
	hold, then
	\begin{equation} \label{Definition_bPsi_star}
	{\bPsi ^*} = \mathop {\arg \max }\limits_{\bPsi  \in \left\{  \bPsi_0^{({\hlambda})},   \bPsi_1^{({\hlambda}) } \right\} } \sum\limits_{i = 0}^1 {{\pi _i}} \lsb \hTEi- \hTLi \rsb,
	\end{equation}
	where $\bPsi_0^{({\hlambda})}$ and $\bPsi_1^{({\hlambda})}$ are defined in (\ref{Define_Psi_01_lambda}).
	
	Moreover, under \emph{Assumptions \ref{Assumption_pq}} and \emph{\ref{Assumption_pq_12}}, the conditions in (C1) and (C2) hold provided that 
	\begin{equation} \label{Define_zeta_star}
	\kappa < \zeta^* \triangleq \min \lcb \zeta_{\hlambda}, \zeta_{\tp,\tq}, \zo \rcb,
	\end{equation}
	where  $\zeta_{\hlambda}$ and $\zo$ are respectively defined in \emph{Lemma \ref{Lemma_Grad_lambda}} and \emph{Lemma \ref{Lemma_Grad_Objective}}, and  $\zeta_{\tp,\tq}$ is defined in (\ref{kappa_smaller_zeta_pq}).
%
\end{theorem}

\begin{IEEEproof}
For the sake of notational simplicity, denote $g_i(\psi_0, \psi_1) \triangleq \hTEi- \hTLi$ and $g(\psi_0, \psi_1) \triangleq \sum_{i=0}^{1}\pi_i [ \hTEi- \hTLi ]$.

If condition (C1) hold, then from (\ref{psi_j_lambda_sup}), we have
\begin{align}
\psi _0^{({\hlambda} )} & = \sup \left\{ {{\psi _0}\left| {\exists {\psi _{1}},{\text{ s.t. }}\bPsi  = {{[{\psi _{0}},{\psi _1}]}} \in {\mathcal{E}}} \right.} \right\}, \\
{\text{and}} \quad \psi _1^{({\hlambda} )} & = \sup \left\{ {{\psi _1}\left| {\exists {\psi _{0}},{\text{ s.t. }}\bPsi  = {{[{\psi _{0}},{\psi _1}]}} \in {\mathcal{E}}} \right.} \right\}.
\end{align}
Hence, according to (C2), 
for any point $\bPsi=[\psi_0, \psi_1]$ in the region  ${\cE}\cap \{ \bPsi: \psi_1 > \psi_0 \ge 0 \}$, 
\begin{equation}
g_i \lp 0, \psi_1^{({\hlambda})} \rp  \ge g_i(0, \psi_1) \ge g_i(\psi_0, \psi_1),
\end{equation}
while in the region ${\cE} \cap \{ \bPsi: \psi_0 > \psi_1 \ge 0 \}$,
\begin{equation}
g_i \lp \psi_0^{({\hlambda})},0 \rp  \ge g_i(\psi_1,0) \ge g_i(\psi_0, \psi_1).
\end{equation}
Therefore, in the region  ${\cE}\cap \{ \bPsi: \psi_1 > \psi_0 \ge 0 \}$, we have
\begin{align} \notag
g\lp 0, \psi_1^{({\hlambda})} \rp & = \sum_{i=0}^{1} \pi_i g_i \lp 0, \psi_1^{({\hlambda})}  \rp \\ \label{Theorem3_temp1}
&= \mathop {\max }\limits_{\bPsi: \psi_1>\psi_0}  \sum_{i=0}^{1}\pi_i \lsb \hTEi- \hTLi \rsb  >0,
\end{align}
where the inequality in (\ref{Theorem3_temp1}) is due to 1) in \emph{Lemma \ref{Lemma_Grad_Objective}}. Similarly, in the region ${\cE} \cap \{ \bPsi: \psi_0 > \psi_1 \ge 0 \}$, we have
\begin{align} \notag
g\lp \psi_0^{({\hlambda})},0  \rp & = \sum_{i=0}^{1} \pi_i g_i \lp \psi_0^{({\hlambda})},0  \rp \\  \label{Theorem3_temp2}
&= \mathop {\max }\limits_{\bPsi: \psi_1<\psi_0}  \sum_{i=0}^{1}\pi_i \lsb \hTEi- \hTLi \rsb >0,
\end{align}
where the inequality in (\ref{Theorem3_temp2}) is also from 1) in \emph{Lemma \ref{Lemma_Grad_Objective}}.
We conclude the proof for  (\ref{Definition_bPsi_star}) by noting the fact that if $\psi_1=\psi_0$, then $\sum_{i=0}^{1}\pi_i [ \hTEi- \hTLi ]=0$ which is smaller than $g( \psi_0^{({\hlambda})},0  )$ and $g( 0, \psi_1^{({\hlambda})})$ from (\ref{Theorem3_temp1}) and (\ref{Theorem3_temp2}).

Furthermore, by employing (\ref{Define_bar_E}), \emph{Lemma \ref{Lemma_Grad_lambda}} and \emph{Lemma \ref{Lemma_Grad_Objective}}, we know that if $\kappa < \zeta^* \triangleq \min \lcb \zeta_{\hlambda}, \zeta_{\tp,\tq}, \zo \rcb$, then the conditions (C1) and (C2) hold.
\end{IEEEproof}

As illustrated in \emph{Theorem \ref{Theorem_Optimal_Solution}}, there exists a constant $\zeta^*$ such that if $\kappa < \zeta^*$ then the optimal solution of the optimization problem in (\ref{Optimal_SE_Problem_no_inf}) can only be either the point $\bPsi_0^{({\hlambda})}$  or $\bPsi_1^{({\hlambda})}$. Thus, the optimization problem in (\ref{Optimal_SE_Problem_no_inf}) can be easily solved, though it is a nonconvex optimization problem. We summarize this procedure in \emph{Algorithm \ref{Algorithm_Solve_Opt}}. The expression of $\hlambda_i(\bPsi)$ can be found in (\ref{lambda_simplified}).  The expressions of $\frac{\partial \lsb \hTEZ- \hTLZ \rsb }{\partial \psi_0}$ and $\frac{\partial \lsb \hTEZ- \hTLZ \rsb }{\partial \psi_1} $ 
can be found in (\ref{for_algorithm_explain_1}) and (\ref{for_algorithm_explain_2}), respectively. The closed-form expressions of the other partial derivatives in condition (C2) can be similarly obtained by following the steps for obtaining (\ref{for_algorithm_explain_1}) and (\ref{for_algorithm_explain_2}) in \emph{Appendix \ref{Proof_Lemma_Grad_Objective}}.
Thus, condition (C2)  can be numerically evaluated. From (\ref{Define_hlambda}), we can obtain $\hlambda_i([0,0])=0$, $i=0,1$. Moreover, by 1) in \emph{Lemma \ref{Lemma_Grad_lambda}}, we know that $\hlambda_i([\psi_0,0])$ and $\hlambda_i([0, \psi_1])$ are strictly increasing with respect to $\psi_0$ and $\psi_1$, respectively. Hence, if $\kappa_i$ satisfies $0\le \kappa_i \le \min\{ \hlambda_i([0, 1]), \hlambda_i([1, 0]) \}$, then by the \emph{Intermediate Value Theorem}, we know that the solution $\psi_j^{({\hlambda})}$ in Step \ref{Step_temp_1} in \emph{Algorithm \ref{Algorithm_Solve_Opt}} exists and is unique. In addition, by the monotonicity of $\hlambda_i([\psi_0,0])$ and $\hlambda_i([0, \psi_1])$,  $\psi_{j, {\hlambda}_i}$ can be easily obtained by numerically searching the point along $\psi_j$-axis which achieves $\hlambda_i(\bPsi) = \kappa_i$  for $i,j=0,1$.
It is worth mentioning that as demonstrated by \emph{Lemma \ref{Lemma_Grad_lambda}} and \emph{Lemma \ref{Lemma_Grad_Objective}}, 
the conditions (C1) and (C2) can always be ensured provided that $\kappa_0$ and $\kappa_1$ are small enough. Note from  (\ref{Define_Psi_01_lambda}), the optimal stochastic encryption strategy is just to flip one type of quantized bit with larger probability and keep the other type of quantized bit unchanged.

\begin{algorithm}[htb]
	\caption{Procedure for computing the optimal $\bPsi^*$ under small $\kappa_i$}
	\begin{algorithmic}[1]
		\STATE \textbf{Input}:  $p$, $q$, and $\kappa_i$;
		\STATE  \textbf{Output}: $\bPsi^*$
		\STATE Evaluate $\psi_j^{({\hlambda})}$: \\ \label{Step_temp_1}
		$[\psi_{0, {\hlambda}_i}, 0] \leftarrow \{\bPsi: {\hlambda}_i(\bPsi) = \kappa_i\} \cap \{\psi_0{\text{--axis}}\}$ \\
		$[0, \psi_{1, {\hlambda}_i}] \leftarrow \{\bPsi:  {\hlambda}_i(\bPsi) = \kappa_i\} \cap \{\psi_1{\text{--axis}}\}$ \\
		$\psi_j^{({\hlambda})} \leftarrow \min\{\psi_{j, {\hlambda}_0}, \psi_{j, {\hlambda}_1} \}$, $j=0,1$
		\IF{the conditions (C1) and (C2) hold} 
		 \STATE ${\bPsi ^*} \leftarrow \mathop {\arg \max }\limits_{\bPsi  \in \left\{  \bPsi_0^{({\hlambda})}=[\psi_0^{({\hlambda})},0],   \bPsi_1^{({\hlambda}) }=[0, \psi_1^{({\hlambda})}]  \right\} } \sum\limits_{i = 0}^1 {{\pi _i}} \lsb \hTEi- \hTLi \rsb $
		\ENDIF 
	\end{algorithmic}
	\label{Algorithm_Solve_Opt}
\end{algorithm}




\begin{figure}
	\begin{minipage}[htb]{0.5\linewidth}
		\centering
\includegraphics[width=0.96\textwidth]{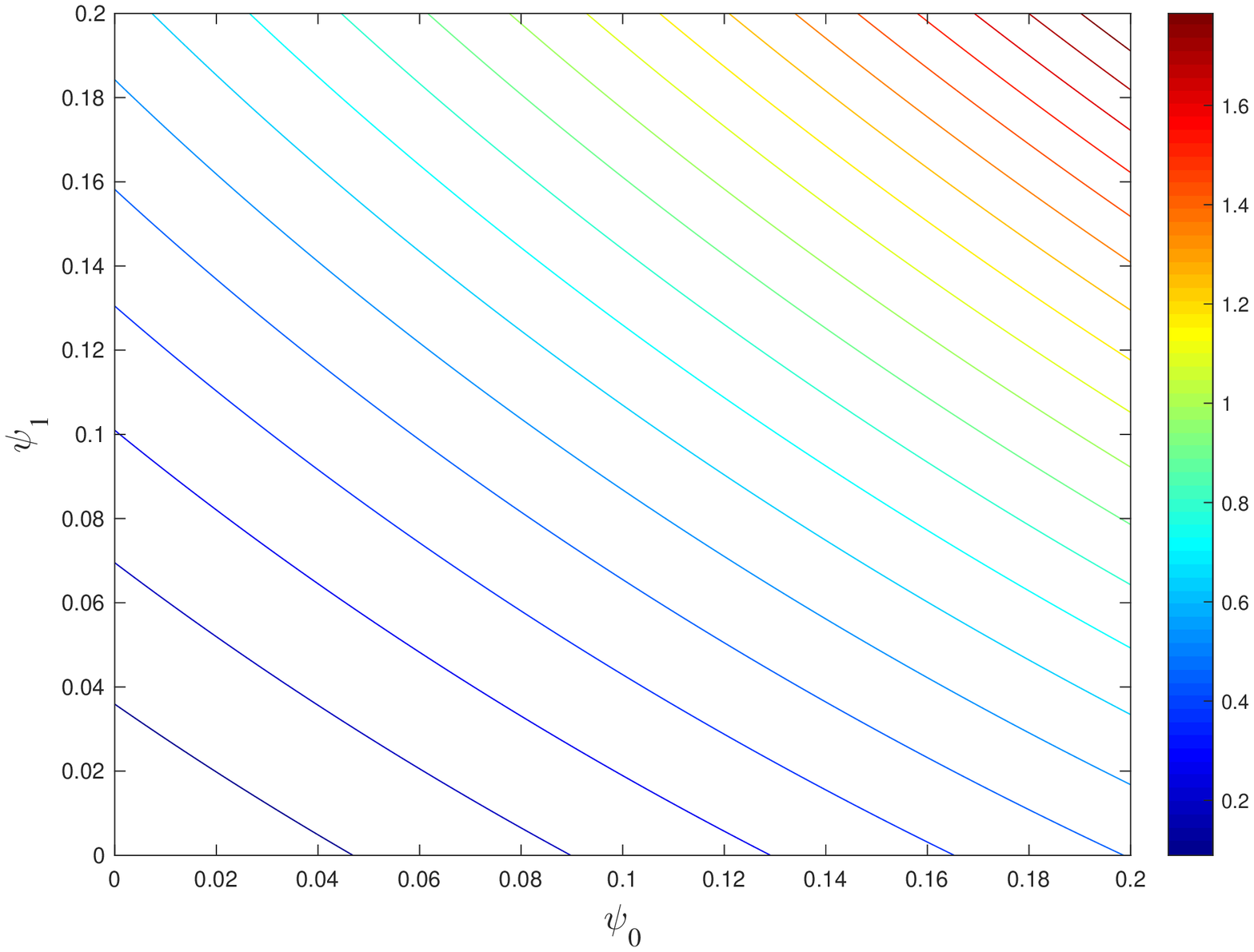}
		\caption{The contour of ${{\hlambda}}_0(\bPsi)$.		
		}	
		\label{Fig_contour_lambda_0}	
	\end{minipage}%
	\begin{minipage}[htb]{0.5\linewidth}
		\centering
		\includegraphics[width=0.96\textwidth]{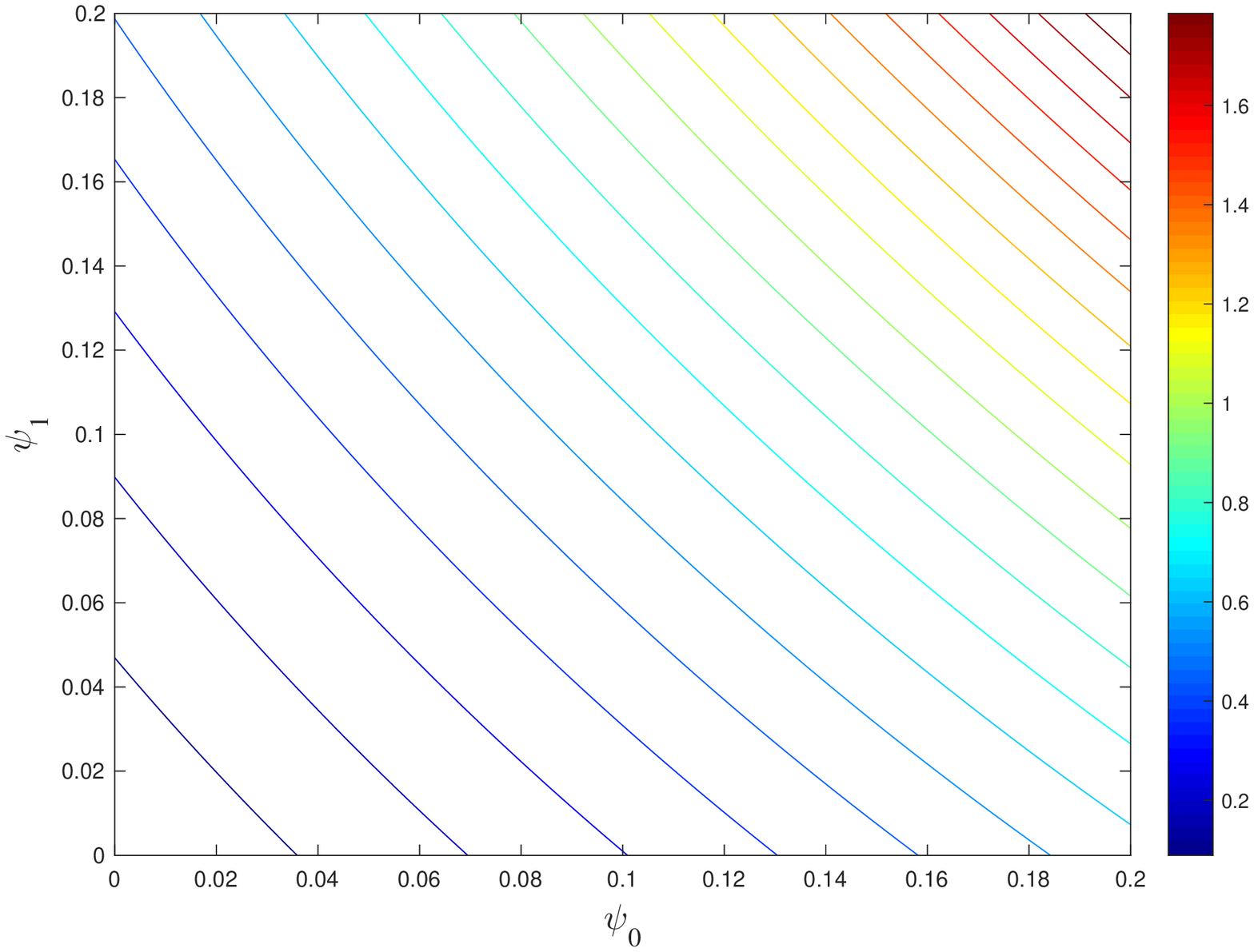}
	\caption{The contour of ${{\hlambda}}_1(\bPsi)$.		
	}	
	\label{Fig_contour_lambda_1}	
	\end{minipage}
\end{figure}
\begin{figure}
	\begin{minipage}[htb]{0.5\linewidth}
		\centering
		\includegraphics[width=0.96\textwidth]{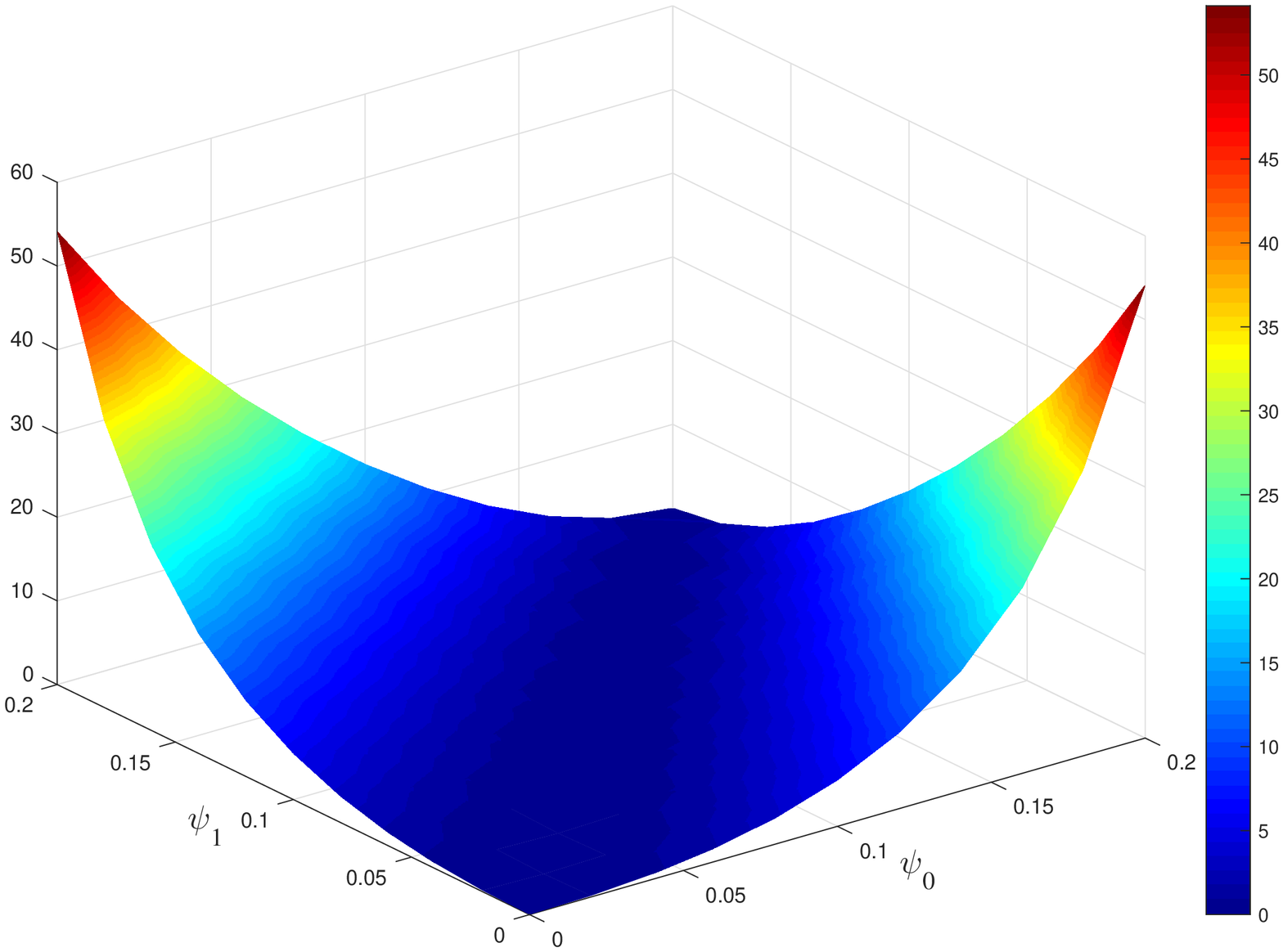}
		\caption{$\hTE(\bPsi) - \hTL(\bPsi)$ versus $\psi_0$ and $\psi_1$.		
		}	
		\label{Fig_E_Obj}	
	\end{minipage}%
	\begin{minipage}[htb]{0.5\linewidth}
		\centering
		\includegraphics[width=0.96\textwidth]{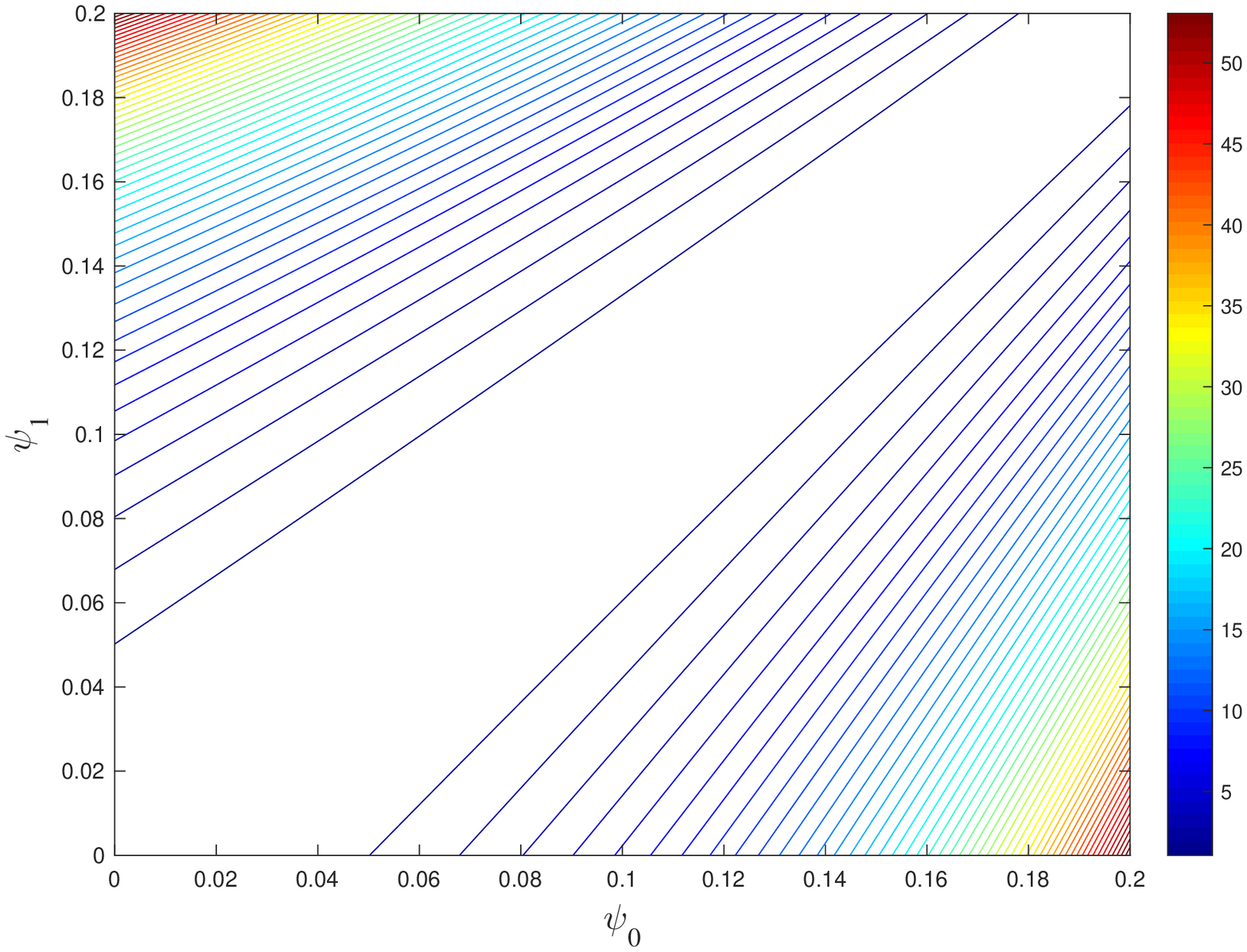}
		\caption{The contour of $\hTE(\bPsi) - \hTL(\bPsi)$.		
		}	
		\label{Fig_contour_Obj}	
	\end{minipage}
\end{figure}

%

%

It is worth mentioning that the constant $\zeta^*$ in (\ref{Define_zeta_star}) does not depend on $\alpha^*$, $\beta^*$ and the stochastic encryption parameter $\bPsi$. 
Moreover, $\bPsi_0^{(\hlambda)}$ and $\bPsi_1^{(\hlambda)}$ do not depend on $\alpha^*$, $\beta^*$, $\pi_0$, and $\pi_1$. The optimal solution $\bPsi^*$ depends on $\alpha^*$, $\beta^*$, $\pi_0$, and $\pi_1$ only through the binary selection in (\ref{Definition_bPsi_star}).
Here, we present an example to illustrate the results in \emph{Lemma \ref{Lemma_Grad_lambda}, Lemma \ref{Lemma_Grad_Objective}} and \emph{Theorem \ref{Theorem_Optimal_Solution}}.

Consider the signal model in (\ref{problem_example}) where $\theta =1$ and the independent noise $w_k^{(n)} \sim {\cal N}(0,1)$. The quantizers employed at the sensors are ${\cQ_n(x) = \bone_{\{x \ge \frac{\theta}{2}\}}}$ for all $n$. In addition, the priors are $\pi_0 = \pi_1 = 0.5$ and the prescribed error probability bounds are $\alpha^*=\beta^*=10^{-6}$. Fig. \ref{Fig_contour_lambda_0} and Fig. \ref{Fig_contour_lambda_1} depict the contours of ${{\hlambda}}_0(\bPsi)$ and ${{\hlambda}}_1(\bPsi)$, respectively. It is seen that the feasible set specified by (\ref{Optimal_SE_Problem_Constraint_Delay_no_inf})--(\ref{Optimal_SE_Problem_Constraint_tq_no_inf}) in this case is nonconvex. Moreover, it is clear that $\nabla_{\bPsi} {{\hlambda}}_i(\bPsi) \succ {\bf 0}$ for $i=0,1$ which corroborates the results in \emph{Lemma \ref{Lemma_Grad_lambda}}. Fig. \ref{Fig_E_Obj} illustrates the objective function $\hTE(\bPsi) - \hTL(\bPsi) \triangleq \pi_0 [\hTEZ - \hTLZ] + \pi_1 [\hTEO - \hTLO]$
in (\ref{Optimal_SE_Problem_Objective_no_inf}) versus the encryption parameters $\psi_0$ and $\psi_1$, and the contours of the objective function $\hTE(\bPsi) - \hTL(\bPsi)$ 
is depicted in Fig. \ref{Fig_contour_Obj}. As expected from \emph{Theorem \ref{Theorem_Optimal_Solution}}, the numerical results in Fig. \ref{Fig_E_Obj} verify that the maximum value of $\hTE(\bPsi) - \hTL(\bPsi)$  can only be attained at either the upper left corner or the lower right corner. Furthermore,  it is seen that the contour curves in Fig. \ref{Fig_contour_Obj} agree with the results in \emph{Lemma \ref{Lemma_Grad_Objective}}. 


\subsection{Simulation Results}
\label{Section_Numerical_Results}

In this subsection, we present a few simulation results to illustrate the performance of the optimal stochastic encryption. 

\begin{figure}
 	\begin{minipage}[htb]{0.5\linewidth}
 		\centering
 		\includegraphics[width=0.96\textwidth]{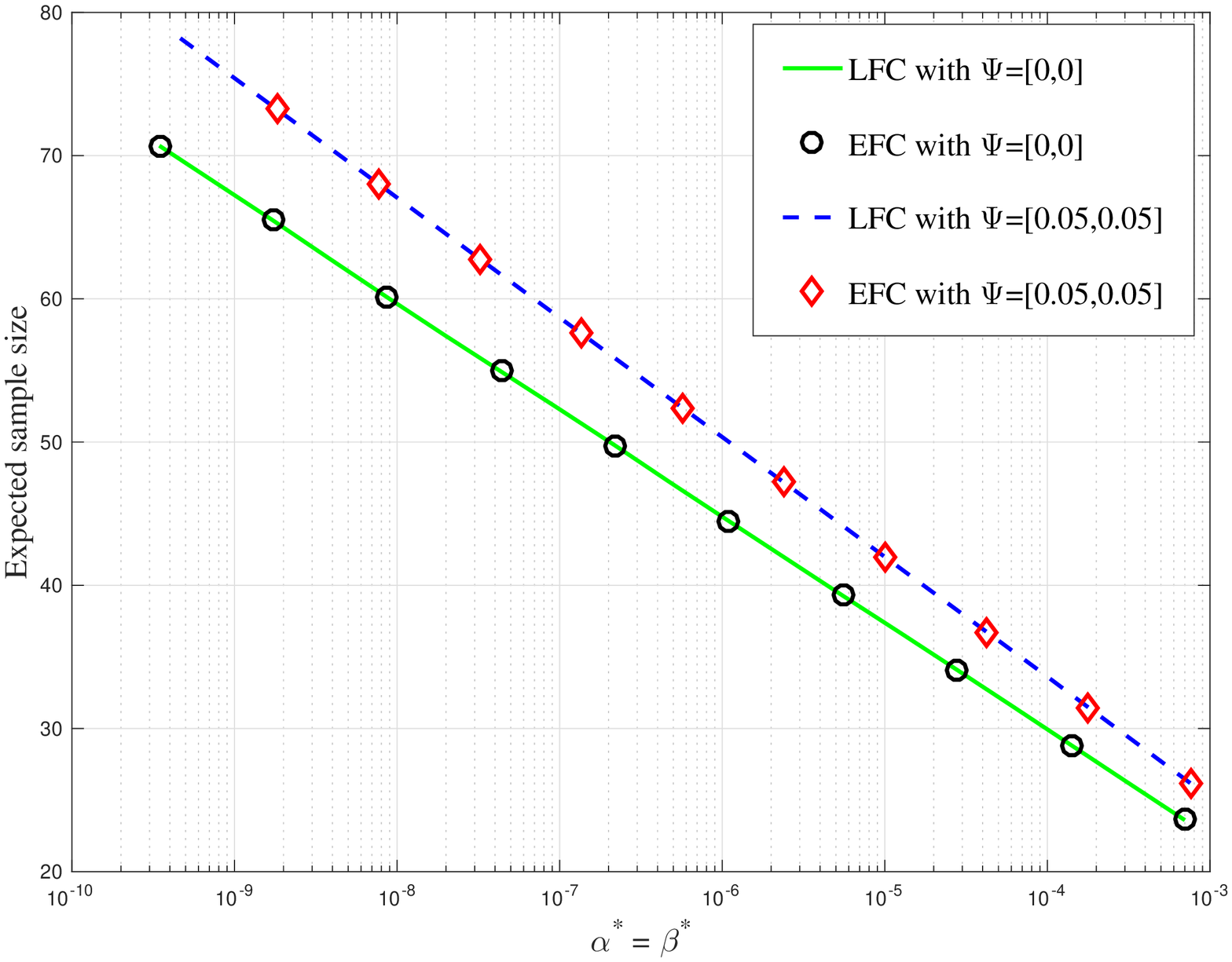}
 		\caption{The expected sample sizes at the LFC and EFC for $\bPsi = [0,0]$ and $\bPsi = [0.05,0.05]$.
 		}	
 		\label{Fig_Simulation_No_and_Symmetric}	
 	\end{minipage}%
 	\begin{minipage}[htb]{0.5\linewidth}
 		\centering
 		\includegraphics[width=0.96\textwidth]{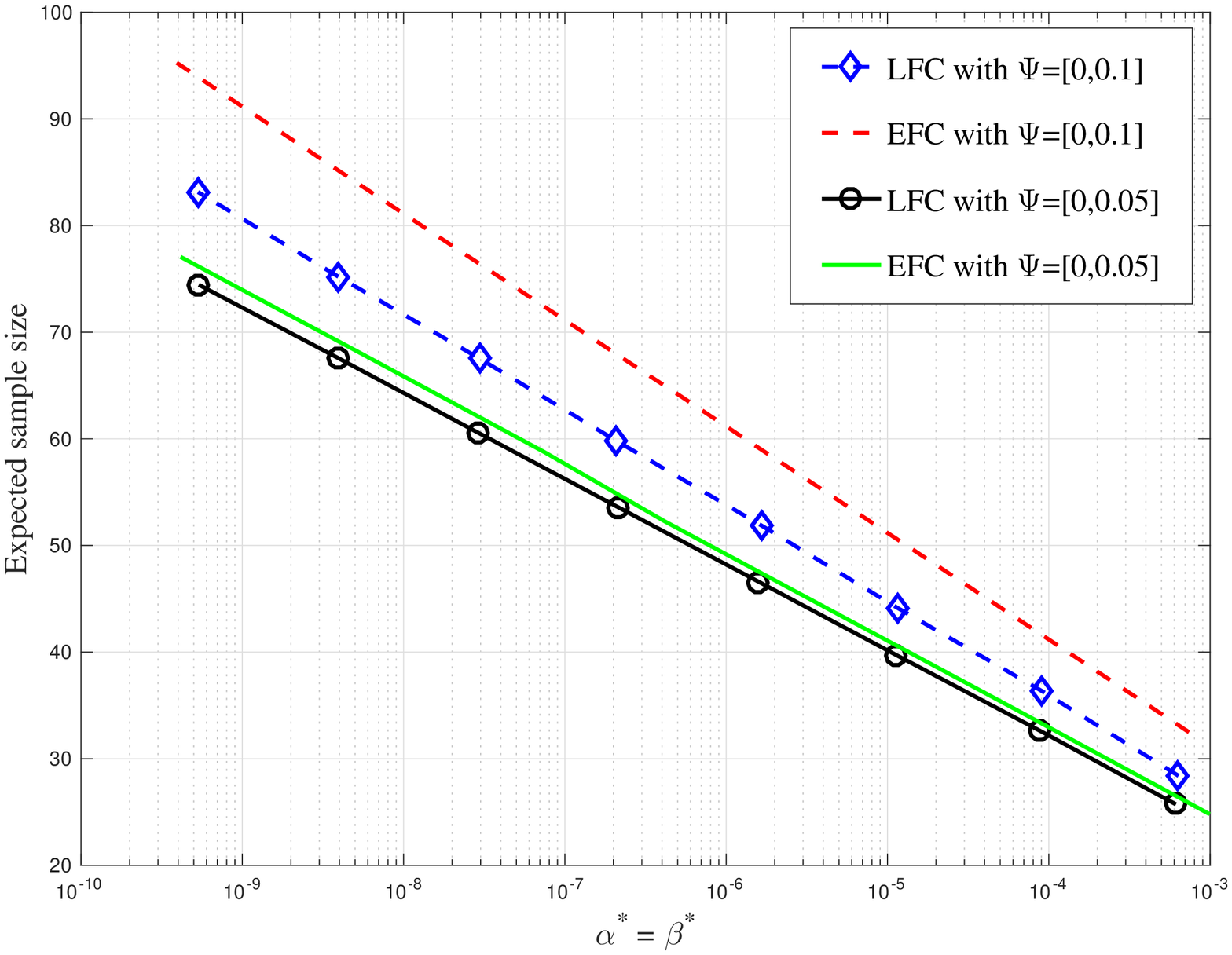}
 		\caption{The expected sample sizes at the LFC and EFC for $\bPsi = [0,0.1]$ and $\bPsi = [0,0.05]$.
 		}	
 		\label{Fig_Simulation_Optimal_and_Nonoptimal}	
 	\end{minipage}
 \end{figure}

The simulation setup considered in this subsection is the same as that for Fig. \ref{Fig_contour_lambda_0}--Fig. \ref{Fig_contour_Obj}. It is seen from Fig. \ref{Fig_contour_lambda_0}--Fig. \ref{Fig_contour_Obj} that if $0\le \psi_0\le0.2$ and $0 \le \psi_1 \le 0.2$, the conditions (C1) and (C2) hold.
 Assume that $\kappa_0 =0.265$ and $\kappa_1 = 0.2077$. By employing Step \ref{Step_temp_1} in \emph{Algorithm \ref{Algorithm_Solve_Opt}}, we can obtain   $\bPsi_0^{({\hlambda})}=[0.08,0]$ and $\bPsi_1^{({\hlambda})}=[0,0.1]$. Moreover, if $\alpha^* = \beta^*$, then by employing (\ref{E_0_T_L_approx})--(\ref{E_1_T_E_approx}), we can obtain $\sum_{i=0}^{1}\pi_i [ \hTE^{(i)}([0,0.1])- \hTL^{(i)}([0,0.1]) ] = 1.756 \sum_{i=0}^{1}\pi_i [ \hTE^{(i)}([0.08,0])- \hTL^{(i)}([0.08,0]) ]$.
Hence, according to \emph{Theorem \ref{Theorem_Optimal_Solution}},  $\bPsi^* = \bPsi_1^{({\hlambda})} = [0,0.1]$ is the optimal solution of (\ref{Optimal_SE_Problem_no_inf}) provided $\alpha^* = \beta^*$. In the following, we compare the performance under different $\bPsi$,
the optimal $\bPsi = [0,0.1]$,  some feasible but not optimal $\bPsi = [0,0.05]$ and $\bPsi = [0.05,0.05]$, and under no stochastic encryption, i.e., $\bPsi = [0,0]$. The average sample sizes over $10^4$ Monte Carlo runs at the LFC and EFC (i.e., $\tbE\{ \TL \} \triangleq \pi_0\tbE_0\{ \TL \} + \pi_1\tbE_1\{ \TL \}$ and  $\tbE\{ \TE \} \triangleq \pi_0\tbE_0\{ \TE \} + \pi_1\tbE_1\{ \TE \}$) versus the prescribed error probability bounds (i.e., $\alpha^*= \beta^*$) are depicted in Fig. \ref{Fig_Simulation_No_and_Symmetric} and Fig. \ref{Fig_Simulation_Optimal_and_Nonoptimal}. As illustrated in Fig. \ref{Fig_Simulation_No_and_Symmetric}, with no stochastic encryption, i.e., $\bPsi=[0,0]$, then the expected sample sizes at the LFC and EFC are the same which agrees with the intuition. In addition, when the symmetric stochastic encryption $\bPsi=[0.05,0.05]$ is employed, the simulation results in Fig. \ref{Fig_Simulation_No_and_Symmetric} verifies that the expected sample sizes at the LFC and EFC are the same as stated by \emph{Theorem \ref{Theorem_Symmetric_Encryption}}. Moreover, the symmetric stochastic encryption $\bPsi=[0.05,0.05]$ causes an increase in the expected sample size at the LFC compared to the case of no stochastic encryption, which verifies that the stochastic encryption incurs performance degradation at the LFC.

In Fig. \ref{Fig_Simulation_Optimal_and_Nonoptimal}, the performances of the optimal stochastic encryption $\bPsi=[0,0.1]$ and the non-optimal stochastic encryption  $\bPsi=[0,0.05]$ are compared. It is seen  that under the optimal stochastic encryption, the difference between the expected sample sizes at the EFC and LFC is significantly  larger than that under the non-optimal one.
However, this is at the price of an increase in the expected sample size at the LFC.
In addition, the slope of $\tbE\{ \TE \}$ is smaller than that of $\tbE\{ \TL \}$, that is, as $\alpha^*=\beta^*$ decreases, $\tbE\{ \TE \}$ grows faster than $\tbE\{ \TL \}$, and therefore, the difference between $\tbE\{ \TE \}$ and $\tbE\{ \TL \}$ becomes larger. 

%
%
%

\section{Conclusions}
\label{Section_Conclusion}

We have investigated sequential detection based on single-bit quantized data and in the presence of eavesdroppers. By employing stochastic encryptions at the sensors, each quantized bit is randomly flipped according to certain probabilities before transmitted to the LFC. 
The LFC knows both the distribution of the quantized data and the flipping probabilities and employs the optimal SPRT; whereas the EFC is unaware of the stochastic encryption and therefore employs a mismatched SPRT. 
We have characterized the expected sample size and the error probabilities of the MSPRT in terms of the detection thresholds. We have shown that when the detection error probabilities are set to be the same at the LFC and EFC, every symmetric stochastic encryption leads to the same expected size at the LFC and EFC. Furthermore, we have provided the asymptotic analysis on the expected sample size in terms of the vanishing error probabilities, and revealed the stark difference from the asymptotic performance of the SPRT with no model mismatch. 
In the asymptotic regime of small detection error probabilities, we have shown that every stochastic encryption degrades the performance of the SPRT at the LFC by increasing the expected sample size, and the expected sample size required at the EFC is no fewer than that is required at the LFC. To this end, symmetric stochastic encryptions are the least favorable ones. Then we have considered the design of the optimal stochastic encryption in the sense of maximizing the difference between the expected sample sizes required at the EFC and LFC. Although this optimization problem is nonconvex, we have shown that if the acceptable tolerance of the increase in the expected sample size at the LFC induced by the stochastic encryption is small enough, the globally optimal stochastic encryption can be analytically obtained. Moreover, the optimal  strategy randomly flips only one type of quantized bits (i.e., $0$ or $1$) and keeps the other type unchanged. 


\appendices

%

\section{Proof of Lemma \ref{Lemma_Grad_lambda}} \label{Proof_Lemma_Grad_lambda}
We first prove $1)$.
From (\ref{lambda_simplified}),  we know that ${\nabla _\bPsi }{\hlambda}_i \left( \bPsi  \right) \succ \boldsymbol{0},  i=0,1$ is equivalent to 
\begin{equation}
\nabla_\bPsi H(\tp,\tq) \prec \boldsymbol{0} \text{ and } \nabla_\bPsi H(\tq,\tp) \prec \boldsymbol{0}.
\end{equation}
By employing (\ref{Define_H}), after some algebra, we can obtain that $\forall j \in \{0,1\}$,
\begin{align} \label{Proof_Lemma_1_temp1}
\frac{{\partial H(\tilde p,\tilde q)}}{{\partial {\psi _j}}} = \frac{{\partial \tilde p}}{{\partial {\psi _j}}}\ln \frac{{\tilde p\left( {1 - \tilde q} \right)}}{{\tilde q\left( {1 - \tilde p} \right)}} - \frac{{\tilde p - \tilde q}}{{\tilde q\left( {1 - \tilde q} \right)}}\frac{{\partial \tilde q}}{{\partial {\psi _j}}}.
\end{align}
Noticing $\tp  > \tq$ and plugging (\ref{Define_tp}) and (\ref{Define_tq}) into (\ref{Proof_Lemma_1_temp1}) yields
\begin{align} \notag
\frac{{\partial H(\tilde p,\tilde q)}}{{\partial {\psi _0}}} & = \left( {1 - p} \right)\ln \frac{{\tilde p\left( {1 - \tilde q} \right)}}{{\tilde q\left( {1 - \tilde p} \right)}} - \left( {1 - q} \right)\frac{{\tilde p - \tilde q}}{{\tilde q\left( {1 - \tilde q} \right)}}\\ \label{Lemma_Grad_H_temp1}
& < \left( {1 - p} \right)\!\!\left[ {\frac{{\tilde p\left( {1 - \tilde q} \right)}}{{\tilde q\left( {1 - \tilde p} \right)}} \!-\! 1} \right] \!-\! \frac{\left( {1 - q} \right)({\tilde p - \tilde q})}{{\tilde q\left( {1 - \tilde q} \right)}}\\ \label{Lemma_Grad_H_temp3}
& = \frac{{\tilde p - \tilde q}}{{\tilde q}}\left( {\frac{{1 - p}}{{1 - \tilde p}} - \frac{{1 - q}}{{1 - \tilde q}}} \right),
\end{align}
\begin{align}  \label{Lemma_Grad_H_psi1}
\text{and} \quad \frac{{\partial H(\tilde p,\tilde q)}}{{\partial {\psi _1}}} & = - p \ln \frac{{\tilde p\left( {1 - \tilde q} \right)}}{{\tilde q\left( {1 - \tilde p} \right)}} + q\frac{{\tilde p - \tilde q}}{{\tilde q\left( {1 - \tilde q} \right)}} \\   \label{Lemma_Grad_H_temp2}
& < - p \lsb 1 - \frac{\tq\lp 1 -\tp \rp}{\tp \lp  1 - \tq \rp} \rsb + q\frac{{\tilde p - \tilde q}}{{\tilde q\left( {1 - \tilde q} \right)}} \\ \label{Lemma_Grad_H_temp4}
& = \frac{{\tilde p - \tilde q}}{{1 - \tilde q}}\left( { - \frac{p}{{\tilde p}} + \frac{q}{{\tilde q}}} \right),
\end{align}
where (\ref{Lemma_Grad_H_temp1}) and (\ref{Lemma_Grad_H_temp2}) are due to the fact that  $\ln x<x-1$ and $\ln x > 1 -1/x$ for all $x>1$.

According to \emph{Assumption \ref{Assumption_pq}}, we know that $p>q$, and therefore, by employing (\ref{Define_tp}) and (\ref{Define_tq}), we can obtain
\begin{equation}  \notag
\frac{p}{{\tilde p}} = \frac{p}{{\left( {1 - {\psi _0} - {\psi _1}} \right)p + {\psi _0}}} \ge \frac{q}{{\left( {1 - {\psi _0} - {\psi _1}} \right)q + {\psi _0}}} = \frac{q}{{\tilde q}},
\end{equation}
\begin{equation} \notag
\text{and    }\frac{{1 - p}}{{1 - \tilde p}} = \frac{1}{{1 - {\psi _0} + {\psi _1}\frac{p}{{1 - p}}}} \le \frac{1}{{1 - {\psi _0} + {\psi _1}\frac{q}{{1 - q}}}} = \frac{{1 - q}}{{1 - \tilde q}},
\end{equation}
which yields
\begin{equation}
\frac{{\partial H(\tilde p,\tilde q)}}{{\partial {\psi _0}}}<0  \text{ and } \frac{{\partial H(\tilde p,\tilde q)}}{{\partial {\psi _1}}} < 0,
\end{equation}
by employing (\ref{Lemma_Grad_H_temp3}) and (\ref{Lemma_Grad_H_temp4}).
Similarly, we can prove $\nabla_\bPsi H(\tq,\tp) \prec \boldsymbol{0}$, and hence, ${\nabla _\bPsi }{\hlambda}_i \left( \bPsi  \right) \succ \boldsymbol{0},  i=0,1.$

Next, we will just prove 
$2)$ for $\psi_1^{({\hlambda})}$, 
and the proof for 
$\psi_0^{({\hlambda})}$ is similar.

It is clear that 
\begin{align} 
\lcb \psi_1 |  {\hlambda}_i \lp \bPsi \rp \le \kappa_i \text{ with }  \psi_{0}=0, \;  i=0,1 \rcb \subset \lcb \psi_1 | \exists \psi_{0}, \text{ s.t. } {\hlambda}_i \lp \bPsi\rp \le \kappa_i,  \;  i=0,1 \rcb.
\end{align}
On the other hand, since we have already proven $\frac{\partial {\hlambda}_i \left( \bPsi  \right) }{\partial \psi_0}> 0$,
we know that if ${\hlambda}_i \lp [\psi_0, \psi_1] \rp \le \kappa_i,  \; i=0,1$, then ${\hlambda}_i \lp [0, \psi_1] \rp < {\hlambda}_i \lp [\psi_0, \psi_1] \rp  \le \kappa_i,  \;  i=0,1$, which yields
\begin{align} 
\lcb \psi_1 |  {\hlambda}_i \lp \bPsi \rp \le \kappa_i \text{ with }  \psi_{0}=0, \;  i=0,1 \rcb  \supset \lcb \psi_1 | \exists \psi_{0}, \text{ s.t. } {\hlambda}_i \lp \bPsi\rp \le \kappa_i,  \;  i=0,1 \rcb.
\end{align} 
Therefore, we can obtain
\begin{align} 
\lcb \psi_1 | \exists \psi_{0}, \text{ s.t. } {\hlambda}_i \lp \bPsi\rp \le \kappa_i,  \;  i=0,1 \rcb  =\lcb \psi_1 |  {\hlambda}_i \lp \bPsi \rp \le \kappa_i \text{ with }  \psi_{0}=0, \;  i=0,1 \rcb,
\end{align} 
which implies that (\ref{Lemma_Grad_H_sup_simplified_1}) is true. 

Furthermore, since we have already proven $\frac{\partial {\hlambda}_i \left( \bPsi  \right) }{\partial \psi_1}> 0$, by the definitions of $\psi_{1, {\hlambda}_0}$ and $\psi_{1, {\hlambda}_1}$, we can obtain
\begin{align} \notag
\sup \lcb \psi_1 |  {\hlambda}_i \lp [0, \psi_1]  \rp \le \kappa_i \;  i=0,1 \rcb  =  \min \lcb \psi_{1, {\hlambda}_0}, \psi_{1, {\hlambda}_1}\rcb,
\end{align}
which completes the proof for (\ref{Lemma_Grad_H_sup_simplified_2}).

In order to prove (\ref{Lemma_Grad_H_sup_simplified_3}), we first define a quantity
\begin{equation}
d_{1,0} \buildrel \Delta \over = \left.  {\frac{{\partial {{\hlambda} _0}( {{{[0,{\psi _1}]}}} )}}{{\partial {\psi _1}}}} \right|_{{\psi _1} = 0} > 0.
\end{equation} 
From (\ref{Define_tp}), (\ref{Define_tq}), (\ref{lambda_simplified}) and (\ref{Lemma_Grad_H_psi1}), it is easy to see that  $\frac{\partial {\hlambda}_0 \left( [0, \psi_1]  \right) }{\partial \psi_1}$ is a continuous function with respect to $\psi_1$.  Thus, there exists $\zeta_{1,0} >0$ such that if $\psi_1 < \zeta_{1,0}$, then 
\begin{equation}
{\frac{{\partial {{\hlambda} _0}( {{{[0,{\psi _1}]}}} )}}{{\partial {\psi _1}}}} \in \left(\frac{d_{1,0}}{2}, \frac{3d_{1,0}}{2} \right).
\end{equation}
Since ${\hlambda}_0 \left( [0, 0]  \right)=0$, we can obtain that if $\kappa_0 < \frac{d_{1,0}}{2} \zeta_{1,0}$, then $\psi_{1, {\hlambda}_0} < \zeta_{1,0}$, and moreover, 
\begin{equation}
\psi_{1, {\hlambda}_0}< \frac{2}{d_{1,0}}\kappa_0 \le \frac{2}{d_{1,0}}\kappa.
\end{equation}

Similarly, there exist two constants $d_{1,1}$ and $\zeta_{1,1}$ such that if $\kappa_1 < \frac{d_{1,1}}{2} \zeta_{1,1}$, then $\psi_{1, {\hlambda}_1}< \frac{2}{d_{1,1}}\kappa$, which implies that if $\kappa < \zeta_{{\hlambda}}^{(1)}$, then
\begin{equation} \label{psi_1_lambda_ub}
\psi_1^{({\hlambda})} = \min \lcb \psi_{1, {\hlambda}_0}, \psi_{1, {\hlambda}_1}\rcb < c_{\hlambda}^{(1)} \kappa,
\end{equation}
where $\zeta_{{\hlambda}}^{(1)}$ and $c_{\hlambda}^{(1)}$ are defined as
\begin{equation} \label{Define_zeta_1}
\zeta_{{\hlambda}}^{(1)} \triangleq \min \lcb \frac{d_{1,0}}{2} \zeta_{1,0}, \frac{d_{1,1}}{2} \zeta_{1,1} \rcb,
\end{equation}  
\begin{equation} \label{Define_c_1}
\text{and    }c_{\hlambda}^{(1)} \triangleq \max \lcb \frac{2}{d_{1,0}},  \frac{2}{d_{1,1}}   \rcb.
\end{equation} 

Analogous to (\ref{Define_zeta_1}) and (\ref{Define_c_1}), there exist two constants $\zeta_{{\hlambda}}^{(0)}$ and $c_{\hlambda}^{(0)}$, such that if $\kappa < \zeta_{{\hlambda}}^{(0)}$, then
\begin{equation} \label{psi_0_lambda_ub}
\psi_0^{({\hlambda})} = \min \lcb \psi_{0, {\hlambda}_0}, \psi_{0, {\hlambda}_1}\rcb < c_{\hlambda}^{(0)} \kappa.
\end{equation}
Therefore, by defining
\begin{equation} \label{proof_Lemma_zeta_lambda}
\zeta_{{\hlambda}}  \triangleq \min \lcb \zeta_{{\hlambda}}^{(0)}, \zeta_{{\hlambda}}^{(1)}  \rcb \text{ and } c_{\hlambda} \triangleq \max \lcb c_{\hlambda}^{(0)}, c_{\hlambda}^{(1)} \rcb,
\end{equation}
we obtain (\ref{Lemma_Grad_H_sup_simplified_3}) from (\ref{psi_1_lambda_ub}) and (\ref{psi_0_lambda_ub}).

\section{Proof of Lemma \ref{Lemma_Grad_Objective}} \label{Proof_Lemma_Grad_Objective}
We first prove $1)$.
By employing (\ref{E_0_T_L_approx}) and (\ref{E_0_T_E_approx}), we can obtain 
\begin{equation} \label{E_T_E_minu_T_L}
\hTEZ - \hTLZ =  f\lp \bPsi \rp \ln \frac{1}{\beta^*},
\end{equation}
\begin{equation} 
\text{with   }f\lp \bPsi \rp \triangleq \bigg\{ \underbrace{{\lp 1 -2\tq \rp } \lsb {\ln \tp - \ln(1-\tp)} \rsb }_{G(\tq,\tp) } \bigg\}^{-1}- \frac{1}{H(\tq,\tp)}.
\end{equation}
Notice that
\begin{align} \notag
G(\tq,\tp) - H(\tq,\tp)  &  = (1 -2\tq) \ln \frac{\tp}{1-\tp}- \tilde q\ln \frac{{\tilde q}}{{\tilde p}} + \left( {1 - \tilde q} \right)\ln \frac{{1 - \tilde q}}{{1 - \tilde p}} \\  \label{Proof_Lemma_2_temp1}
& = -\lsb  (1-\tilde q)\ln \frac{{1 - \tilde q}}{{\tilde p}} + { \tilde q}\ln \frac{{ \tilde q}}{{1 - \tilde p}}  \rsb,
\end{align}
where the inside of the bracket in (\ref{Proof_Lemma_2_temp1}) is the  KL divergence of two Bernoulli distributions, and therefore,
\begin{equation} \label{G_smaller_H}
G(\tq,\tp) \le H(\tq,\tp),
\end{equation}
with equality if and only if $\tp + \tq=1$. Hence, from (\ref{Define_tp}) and (\ref{Define_tq}), we know that
\begin{equation}
\hTEZ - \hTLZ \ge 0
\end{equation}
with equality if and only if $\psi_0=\psi_1$, which proves $1)$ 
for $\hTEZ - \hTLZ $. The proof for $\hTEO - \hTLO  \ge 0$ with equality if and only if $\psi_0 =\psi_1$ is similar.

Next, we consider $2)$. We first prove 
(\ref{Grad_Objective_psi1_bigger_psi0}) 
for $\hTEZ - \hTLZ$  in the region ${\bar{\cE}}\cap \{ \bPsi: \psi_1 > \psi_0 \ge 0 \}$.
Denote $\delta\triangleq \psi_1 - \psi_0 > 0$.
Under \emph{Assumption \ref{Assumption_pq}}, by employing (\ref{Define_tp}) and (\ref{Define_tq}), we can obtain
\begin{equation} \label{tp_tq_delta}
\tp = 1 - \tq - \delta.
\end{equation}

By taking partial derivative of $f\lp \bPsi \rp$ with respect to $\psi_j$, after some algebra, we can obtain
\begin{align} \notag
 \frac{{\partial f\left( \bPsi  \right)}}{{\partial {\psi _j}}}   =  & {\underbrace {\left[ \frac{1}{{H{{\left( {\tilde q,{\rm{ }}\tilde p} \right)}^2}}}\frac{{\tilde p - \tilde q}}{{\tilde p\left( {1 - \tilde p} \right)}} - \frac{1}{{G{{\left( {\tilde q,\tilde p} \right)}^2}}}\frac{{1 - 2\tilde q}}{{\tilde p\left( {1 - \tilde p} \right)}}\right]}_{{Y_1(\tq, \delta)}}} \frac{{\partial \tilde p}}{{\partial {\psi _j}}}  \\ \label{Partial_f}
&  \! + \!  {\underbrace { \left[  \frac{2}{{G{{\left( { \tilde q,\tilde p} \right)}^2}}}\ln \frac{{\tilde p}}{{1 - \tilde p}} - \frac{1}{{H{{\left( {\tilde q,{\rm{ }}\tilde p} \right)}^2}}}\ln \frac{{\tilde p\left( {1 - \tilde q} \right)}}{{\tilde q\left( {1 - \tilde p} \right)}}  \right] }_{{Y_2(\tq, \delta)}}} \frac{{\partial \tilde q}}{{\partial {\psi _j}}}.
\end{align}

From (\ref{tp_tq_delta}), we know that $\tp < 1-\tq$ and $1-2\tq > \tp - \tq$, and hence
\begin{equation} \label{Y1_smaller_0}
Y_1(\tq, \delta) < 0,
\end{equation}
since $G(\tq,\tp) < H(\tq,\tp)$ as illustrated in (\ref{G_smaller_H}).

On the other hand, ${Y_2}\left( {\tilde q,\delta } \right)$ can be rewritten as
\begin{align} \label{Y2_Z1_Z2}
{Y_2}\left( {\tilde q,\delta } \right)  = \underbrace{\frac{1}{{{{(1 - 2\tilde q)}^2}H{{\left( {\tilde q,{\rm{ }}\tilde p} \right)}^2}\ln \frac{{\tilde p}}{{1 - \tilde p}}}}}_{Z_0(\tq,\delta)} \Bigg[ \underbrace {2H{{\left( {\tilde q,{\rm{ }}\tilde p} \right)}^2}}_{{Z_1}\left( {\tilde q,\delta } \right)} \Bigg.  \Bigg. - \underbrace {{{(1 - 2\tilde q)}^2}\ln \frac{{\tilde p}}{{1 - \tilde p}}\ln \frac{{\tilde p\left( {1 - \tilde q} \right)}}{{\tilde q\left( {1 - \tilde p} \right)}}}_{{Z_2}\left( {\tilde q,\delta } \right)} \Bigg],
\end{align}
which implies
\begin{equation} \label{Z_1_minus_Z_2_equal_0}
{Y_2}\left( {\tilde q, 0 } \right)  = {Z_1}\left( {\tilde q,0 } \right) - {Z_2}\left( {\tilde q, 0 } \right) =0.
\end{equation}
Furthermore, by taking partial derivative of ${Z_1}\left( {\tilde q, \delta } \right)$ and ${Z_2}\left( {\tilde q, \delta} \right)$ with respect to $\delta$, we can obtain
\begin{align} 
\frac{{\partial {Z_1}\left( {\tilde q,\delta } \right)}}{{\partial \delta }} = 2{H\left( {\tilde q, 1 - \tilde q - \delta } \right)}\frac{\partial {H\left( {\tilde q, 1 - \tilde q - \delta } \right)}}{\partial \delta} = \frac{{4H\left( {\tilde q, 1 - \tilde q - \delta } \right)}}{{(1 - \tilde q - \delta )\left( {\tilde q + \delta } \right)}}\left( {2\tilde q + \delta  - 1} \right) <0,
\end{align}
since $2\tilde q + \delta  - 1 = \tq -\tp <0$ according to \emph{Assumption \ref{Assumption_pq_12}}, and
\begin{align} \notag
\frac{{\partial {Z_2}\left( {\tilde q,\delta } \right)}}{{\partial \delta }} &  = {(1 - 2\tilde q)^2}\left[ {\ln \frac{{\tilde p\left( {1 - \tilde q} \right)}}{{\tilde q\left( {1 - \tilde p} \right)}}\frac{\partial \ln \frac{{\tilde p}}{{1 - \tilde p}}}{{\partial \delta }} + \ln \frac{{\tilde p}}{{1 - \tilde p}}\frac{\partial \ln \frac{{\tilde p\left( {1 - \tilde q} \right)}}{{\tilde q\left( {1 - \tilde p} \right)}}}{{\partial \delta }} } \right]\\
& =   \frac{{{{-(1 - 2\tilde q)}^2}}}{{(1 - \tilde q - \delta )\left( {\tilde q + \delta } \right)}}\left( {2\ln \frac{{1 - \tilde q - \delta }}{{\tilde q + \delta }} + \ln \frac{{1 - \tilde q}}{{\tilde q}}} \right),
\end{align}
which yields
\begin{align}   \label{differential_Z1_minus_Z2}
d_Z\left( {\tilde q,\delta } \right)  \triangleq  \frac{{\partial \lsb {Z_1}\left( {\tilde q,\delta } \right) - {Z_2}\left( {\tilde q,\delta } \right) \rsb }}{{\partial \delta }}  =\frac{{{{Z_d(\tq,\delta)}}}}{{(1 - \tilde q - \delta )\left( {\tilde q + \delta } \right)}},
\end{align}
\begin{align} 
\text{with      } {Z_d}\left( {\tilde q,\delta } \right)  \buildrel \Delta \over = -4\left( {1 - 2\tilde q - \delta } \right)H\left( {\tilde q,{\rm{ }}1 - \tilde q - \delta } \right) + {(1 - 2\tilde q)^2}\left( {2\ln \frac{{1 - \tilde q - \delta }}{{\tilde q + \delta }} + \ln \frac{{1 - \tilde q}}{{\tilde q}}} \right).
\end{align}
Note that
\begin{equation} 
{Z_d}\left( {\tilde q, 0 } \right) = -{(1 - 2\tilde q)^2} \ln \frac{1- \tq}{\tq}  <0,
\end{equation}
since $\tq<0.5$ according to \emph{Assumption \ref{Assumption_pq_12}}. Hence, from (\ref{differential_Z1_minus_Z2}), we can obtain that for all $\tq \in \lp 0,\frac{1}{2} \rp$,
\begin{equation} \label{differential_Z1_minus_Z2_at_0}
d_Z\left( {\tilde q,0 } \right)  = - \frac{{{{{(1 - 2\tilde q)^2}   }}}}{{(1 - \tilde q  ) {\tilde q  } }} \ln \frac{1- \tq}{\tq} <  0 
\end{equation}
since $1 - \tilde q >0$ and $\tilde q >0$.

Define a constant $\zeta_{0,1}^{(0)}$
\begin{equation} \label{Define_zeta_o_0}
\zeta_{0,1}^{(0)}\triangleq \frac{1-2q}{4c_{\hlambda}(1-q)},
\end{equation}
where $c_{\hlambda}$ is the constant in (\ref{Lemma_Grad_H_sup_simplified_3}).

By employing \emph{Lemma \ref{Lemma_Grad_lambda}}, we know that if
\begin{equation}
\kappa < \min \lcb\zeta_{\hlambda}, \zeta_{0,1}^{(0)} \rcb,
\end{equation}
then $\psi_0<c_{\hlambda} \kappa$ and $\psi_1<c_{\hlambda} \kappa$, and therefore, from (\ref{Define_tq}), we can obtain
\begin{align} \label{tq_compact}
\tq & =  q + (1-q)\psi_0 - \psi_1q \in [q-c_{\hlambda}\kappa q, q +c_{\hlambda} \kappa (1-q)]  \\ \label{tq_range}
& \subset \lp q-\frac{1-2q}{4(1-q)}q, \frac{1}{4}+\frac{1}{2}q \rp \subset \lp 0,\frac{1}{2} \rp,
\end{align}
where (\ref{tq_range}) is due to (\ref{Define_zeta_o_0}) and the fact that $0 < q<0.5$. From (\ref{differential_Z1_minus_Z2_at_0}), we know that $d_Z\left( {\tilde q,0 } \right) $ is a continuous function with respect to $\tq$ over $\tq \in \lp 0,\frac{1}{2} \rp$, and hence, can achieve its maximum $d_Z^* <0$ over the closed set $[q-c_{\hlambda}\kappa q, q +c_{\hlambda} \kappa (1-q)]$, that is,
\begin{equation} \label{d_Z_maximum}
d_Z\left( {\tilde q,0 } \right) \le d_Z^* <0, \forall \tq \in [q-c_{\hlambda}\kappa q, q +c_{\hlambda} \kappa (1-q)].
\end{equation}

Notice from (\ref{differential_Z1_minus_Z2}) that 
$d_Z\left( {\tilde q,\delta } \right)$ is continuous with respect to $(\tq, \delta)$. Thus, from (\ref{differential_Z1_minus_Z2_at_0}), we know that $\forall \tq\in [q-c_{\hlambda}\kappa q, q +c_{\hlambda} \kappa (1-q)]$, there exists $\varepsilon_{\tq} >0$ such that if 
\begin{equation}
(\tq, \delta) \in (\tq - \varepsilon_{\tq}, \tq + \varepsilon_{\tq}) \times [0, \varepsilon_{\tq})
\end{equation}
then
\begin{equation} \label{Lemma_3_temp1}
d_Z\left( {\tilde q,\delta } \right) \in \lp \frac{3}{2}d_Z\left( {\tilde q,0 } \right), \frac{1}{2}d_Z\left( {\tilde q,0 } \right) \rp.
\end{equation}
Noting that 
\begin{equation} \notag
[q-c_{\hlambda}\kappa q, q +c_{\hlambda} \kappa (1-q)] \! \subset \!\!\!\!\!\! \mathop { \bigcup  }\limits_{\tilde q \in [q - c_{\hlambda}\kappa q,q + c_{\hlambda}\kappa (1 - q)]} \!\!\!\!\!\! \left( {\tilde q - {\varepsilon _{\tilde q}},\tilde q + {\varepsilon _{\tilde q}}} \right), 
\end{equation}
and the set $[q-c_{\hlambda}\kappa q, q +c_{\hlambda} \kappa (1-q)]$ is compact, we know that there exist $\{\tq_1, \tq_2,...,\tq_M\} \subset [q-c_{\hlambda}\kappa q, q +c_{\hlambda} \kappa (1-q)]$ such that
\begin{equation}
[q-c_{\hlambda}\kappa q, q +c_{\hlambda} \kappa (1-q)]  \subset \mathop  \bigcup \limits_{i = 1}^M \left( {{{\tilde q}_i} - {\varepsilon _{{{\tilde q}_i}}},{{\tilde q}_i} + {\varepsilon _{{{\tilde q}_i}}}} \right).
\end{equation}
By defining $\varepsilon_Z \triangleq \min \{\varepsilon_{\tq_1}, \varepsilon_{\tq_2},...,\varepsilon_{\tq_M}\}$, we can obtain from (\ref{Lemma_3_temp1}) that if $\delta \in  [0, \varepsilon_{Z})$, then $\forall \tq\in [q-c_{\hlambda}\kappa q, q +c_{\hlambda} \kappa (1-q)]$,
\begin{equation} \label{d_Z_ub}
d_Z\left( {\tilde q,\delta } \right) < \frac{1}{2}d_Z\left( {\tilde q,0 } \right) \le \frac{1}{2}d_Z^*<0.
\end{equation}

Let $\zeta_{0,1}^{(1)}$ denote a constant
\begin{equation}
\zeta_{0,1}^{(1)} \triangleq \frac{\varepsilon_Z}{c_{\hlambda}}.
\end{equation}
If $\kappa < \min \{\zeta_{{\hlambda}}, \zeta_{0,1}^{(0)}, \zeta_{0,1}^{(1)}\}$, then by employing \emph{Lemma \ref{Lemma_Grad_lambda}}, we can obtain that $\forall \bPsi \in {\bar{\cE}}\cap \{ \bPsi: \psi_1 > \psi_0 \ge 0 \}$,
\begin{equation}
\psi_0 \in [0, \varepsilon_Z) \text{   and   } \psi_1 \in [0, \varepsilon_Z),
\end{equation}
and hence, $\delta = \psi_1 - \psi_0 \in (0, \varepsilon_Z)$, which implies that $\forall \tq \in [q-c_{\hlambda}\kappa q, q +c_{\hlambda} \kappa (1-q)]$ and $\forall \delta \in (0, \varepsilon_Z)$, by employing Taylor's theorem, there exists a $\tilde{\delta} \in (0, \delta)$ such that
\begin{align} 
{Z_1}\left( {\tilde q, \delta} \right) - {Z_2}\left( {\tilde q, \delta } \right)  = \lsb {Z_1}\left( {\tilde q, 0} \right) - {Z_2}\left( {\tilde q, 0 } \right) \rsb  +   d_Z( {\tilde q,\tilde{\delta}  } )  \delta = d_Z( {\tilde q,\tilde{\delta}  } )  \delta  \le  \frac{1}{2}d_Z^* \delta <0
\end{align}
from (\ref{Z_1_minus_Z_2_equal_0}) and (\ref{d_Z_ub}). As a result, by employing (\ref{Y2_Z1_Z2}), we can obtain
\begin{equation} \label{Y_2_smaller_0}
Y_2\left( {\tilde q, \delta} \right) = Z_0\left( {\tilde q, \delta} \right)\lp {Z_1}\left( {\tilde q, \delta} \right) - {Z_2}\left( {\tilde q, \delta } \right) \rp <0,
\end{equation}
since $Z_0\left( {\tilde q, \delta} \right)>0$ which is a consequence of  $\tp>0.5>\tq$ according to \emph{Assumption \ref{Assumption_pq_12}}. 

Furthermore, from (\ref{Define_tp}) and (\ref{Define_tq}), we can obtain
\begin{equation} \label{Lemma_3_temp_2}
\frac{\partial \tp}{\psi_0} = -\frac{\partial \tq}{\psi_1}  = 1-p \text{ and } \frac{\partial \tq}{\psi_0} = -\frac{\partial \tp}{\psi_1}  = p.
\end{equation}
Therefore, by defining
\begin{equation}
\zeta_{0,1} = \min \{\zeta_{{\hlambda}}, \zeta_{0,1}^{(0)}, \zeta_{0,1}^{(1)}\},
\end{equation}
and by employing (\ref{E_T_E_minu_T_L}), (\ref{Partial_f}), (\ref{Y1_smaller_0}), (\ref{Y_2_smaller_0}) and (\ref{Lemma_3_temp_2}), we know that if $\kappa < \zeta_{0,1}$, then in the region ${\bar{\cE}}\cap \{ \bPsi: \psi_1 > \psi_0 \ge 0 \}$,
\begin{equation} \label{for_algorithm_explain_1}
\frac{\partial \lsb \hTEZ - \hTLZ \rsb  } {\partial \psi_0} = \lsb (1-p)Y_1(\tq, \delta) + p Y_2(\tq, \delta) \rsb \ln \frac{1}{\beta^*} <0,
\end{equation}
\begin{equation}  \label{for_algorithm_explain_2}
\text{and} \quad \frac{\partial \lsb \hTEZ - \hTLZ \rsb } {\partial \psi_1} = -\lsb pY_1(\tq, \delta) + (1-p) Y_2(\tq, \delta) \rsb \ln \frac{1}{\beta^*} >0,
\end{equation}
since $\beta^*<1$, which complete the proof of (\ref{Grad_Objective_psi1_bigger_psi0}) for $\hTEZ - \hTLZ$ .

By following a similar approach as above, we can show that in the region ${\bar{\cE}}\cap \{ \bPsi: 0 \le \psi_1 < \psi_0 \}$, there exists a constant $\zeta_{0,0} $ such that if $\kappa < \zeta_{0,0}$, then (\ref{Grad_Objective_psi1_smaller_psi0})  for $\hTEZ - \hTLZ$  is true. Let $\zo^{(0)} \triangleq \min \{  \zeta_{0,0}, \zeta_{0,1} \}$.

Furthermore,
we can similarly prove that there exists a constant $\zo^{(1)}  $ such that if $\kappa < \zo^{(1)} $, then the corresponding results in (\ref{Grad_Objective_psi1_bigger_psi0}) and (\ref{Grad_Objective_psi1_smaller_psi0}) for $\hTEO - \hTLO$ are true. 
For the sake of brevity, the details of these proofs are omitted.
Finally, we can conclude the proof for $2)$ by defining
\begin{equation}
\zo \triangleq \min \lcb \zo^{(0)}, \zo^{(1)} \rcb.
\end{equation} 

\bibliographystyle{IEEEtran}
\bibliography{IEEEabrv,StochasticEncryption}

\end{document}